\documentclass[aps,prb,reprint,amsmath,amssymb,longbibliography,floatfix]{revtex4-2}
\usepackage[T1]{fontenc}
\usepackage{lmodern}
\usepackage{graphicx,booktabs,bm,mathtools,microtype,xcolor}
\usepackage[colorlinks=true,linkcolor={blue!45!black},citecolor={blue!45!black},urlcolor={blue!45!black}]{hyperref}
\usepackage{placeins}
\newcommand{\Js}{J_{\sigma}}
\newcommand{\absI}{I^{\mathrm{abs}}}

\newcommand{\QuenchSchematicCaption}{Local spin quench and intensity matching (schematic). (a) The spin insertion reverses one nearest-neighbor bond, toggles both adjacent fluxes, and reverses four next-nearest-neighbor hoppings (dashed). Both responses use this changed Hamiltonian. (b) On an eligible odd source, the full unit-mass spectrum contains one-particle weight $w_1$ and higher-sector weight $1-w_1$. The approximation $P$ rescales the same one-particle lines to unit mass; $h$ is not separately normalized. Even-source responses are shared. (c) The depth is the mean intensity at the two flanking frequencies minus the center. (d) Open squares denote the matched one-particle response $P$ and filled circles the full response $S$. Positive $D=d(P)-d(S)$ means shallowing; negative $D$ means deepening. A brighter center can accompany still larger gains at the flanks. All bar partitions and triplets are illustrative, not representative spectra or pointwise medians. Figure~\ref{fig:reversal} supplies the quantitative result. ChatGPT and Codex assisted with plot and figure preparation.}
\newcommand{\ReversalCaption}{The paired correction changes sign across finite tori. (a) Median of $D=d(P)-d(S)$ conditioned on eligible odd sources. Positive means shallowing when higher sectors are restored; negative means deepening. The 24-site diamonds use exact conditional orbit probabilities, with numerical uncertainty smaller than the symbols. Bars on the sampled cohorts are simultaneous population-median enclosures. The first six positions form the homothetic series; the last is a rectangular shape comparison. (b) Numbers of eligible 6,144-site backgrounds with central enhancement, deepening, or both. Each interval runs from the definite to the conservative possible count; the dotted line marks half of the 170 odd draws. The 150 shared even sources have zero correction. Together with the odd-source sign counts, they fix the all-source sample median at zero (Sec.~S2). The exact 24-site all-source median is also zero. ChatGPT and Codex assisted with plot and figure preparation.}
\newcommand{\AbsoluteCaption}{Mean spectra on the 24-site and 6,144-site tori in absolute spin units. Open squares show the intensity-matched approximation $P/4$ and filled circles the full response $S/4$, using peak-height-one Lorentzian filters. For each spectrum, its depth is the vertical distance from the midpoint of the line joining the two flanks down to its central reading [Fig.~\ref{fig:schematic}(c,d)]. The top and bottom rows use $\gamma/\Js=0.05$ and $0.10$. Frequencies are signed transfers. Left: exact 24-site orbit average conditioned on odd eligibility. Middle: mean of the 170 eligible odd backgrounds in the realized 6,144-site cohort. Right: exact 24-site average over all backgrounds, including the shared even responses. Points mark numerical-enclosure midpoints, with small horizontal offsets for legibility; most whiskers are smaller than the symbols. Each annotation reports $10^3\overline D_{\rm abs}$ from the tighter bounds in Sec.~S3, rather than the separately rounded Table~S1 ordinates. The odd-source mean correction is positive on 24 sites and negative on 6,144 sites at both resolutions. In panel (b) the intensity rises at the center and rises more at both flanks. Table~S5 separates the added and removed weights. Total unfiltered local $S^z$ intensity is $1/4$. ChatGPT and Codex assisted with plot and figure preparation.}

\newcommand{\BenchmarkCaption}{Diagnostic prediction fixed before the physical comparison at $b/J=0.50$. The frozen sector mixture is not a ground-state ensemble and retains approximately 13.72\% unbroadened negative-transfer mass. Squares and circles show the intensity-matched and full predictions averaged with the exact 24-site gauge-orbit probabilities. Only odd sources are truncated in the intensity-matched approximation. Both curves include the full even-source response. Triangles are the physical local $S^z_0$ response on the same torus, site, energy scale, and $\gamma/J_\sigma=0.05$ filter. Each point is a filtered absolute spin intensity, including the elastic contribution. The underlying unfiltered local measure has mass $1/4$. The largest reduction in squared error occurs on the left. Both model responses place the right shoulder above the center, whereas the physical response places it below. Whiskers show numerical enclosures. Horizontal offsets separate the three measurements at each energy. ChatGPT and Codex assisted with plot and figure preparation.}

\hypersetup{pdftitle={Finite-size reversal of a multiparticle correction to local spin spectra in a flux-disordered Kitaev honeycomb model},pdfauthor={David C. Ronquillo}}
\begin{document}

\title{Finite-size reversal of a multiparticle correction to local spin spectra in a flux-disordered Kitaev honeycomb model}
\author{David C. Ronquillo}
\email{david.ronquil@gmail.com}
\noaffiliation
\date{November 21, 2026}

\begin{abstract}
A spin flip in the Kitaev honeycomb model changes the flux background and therefore the Hamiltonian seen by the matter fermions. Can a one-particle approximation with the correct total spin intensity nevertheless misestimate the depth of a depression in the local spin spectrum? We test this question in a frozen, flux-disordered ensemble by retaining the actual one-particle transition energies and relative spectral weights after the spin flip, then rescaling those weights to match the full intensity. We measure the depression by subtracting the central intensity from the average at two flanking frequencies. For spin-created states with odd final quasiparticle number and nonzero one-particle weight, restoring multiparticle weight shallows the depression on 24 sites and deepens it on both 1,536-site shapes and the 3,456- and 6,144-site tori, as judged by the median paired correction at each of two energy resolutions. For these states, the one-particle approximation overestimates the mean dip depth by about 16--18\% on 24 sites and underestimates it by 33--56\% in the 6,144-site sample. Extrapolating from 24 sites would therefore give the wrong sign for this correction on those four larger tori under the same disorder prescription.
\end{abstract}

\maketitle

\section{Introduction}\label{sec:intro}

Spin spectroscopy probes more than the density of matter-fermion states in a Kitaev spin liquid. In the spin-$1/2$ honeycomb model, each spin component is represented by the product of a matter Majorana and a gauge Majorana. The gauge variables set bond signs, whose products around hexagons are the plaquette fluxes~\cite{Kitaev2006,Baskaran2007}. For a fixed set of signs the matter Hamiltonian is quadratic. Acting with a local spin changes the matter state and reverses one bond sign, toggling the two adjacent fluxes. The matter state then propagates with this changed Hamiltonian until the second spin insertion. This is the local quench in the spin correlation function~\cite{Knolle2014,Knolle2015}. In a frozen-flux calculation the fluxes have no autonomous dynamics between insertions; the spin operator still changes the background. Expanding the prepared matter state in the new eigenbasis admits several quasiparticles, with intensities fixed by squared spin-transition amplitudes.

How much of this response can a one-particle description retain? In a clean isotropic example with vanishing three-spin coupling, Knolle \emph{et al.} found about 97.5\% of the integrated zero-momentum $zz$ structure factor in the one-particle channel~\cite{Knolle2015}. Two local spin spectra can have the same integrated intensity and distribute it differently across frequencies. We test how well the intensity-matched one-particle approximation reproduces a spectral depression by comparing the central intensity with the average at two flanking frequencies. On each background with nonzero one-particle weight we keep the energies and relative spectral weights of the actual one-particle transitions after the spin flip, and rescale their weights to the full local spin sum rule. Restoring the omitted sectors can then either fill the depression or deepen its contrast by adding still more intensity on its flanks. To test whether the small cluster predicts the direction of this correction on larger tori, we compare the same frozen-flux model across sizes. On 24 sites we can enumerate all flux backgrounds up to gauge equivalence and sum over the allowed final occupations. On larger tori we use the quadratic matter Hamiltonian to evaluate transition weights on sampled backgrounds and bound the contribution of unresolved states. A separate calculation in Sec.~\ref{sec:field} tests how closely the frozen model reproduces the local spin response of the interacting 24-spin magnet in a field.

\begin{figure*}[t]
\includegraphics[width=.98\textwidth]{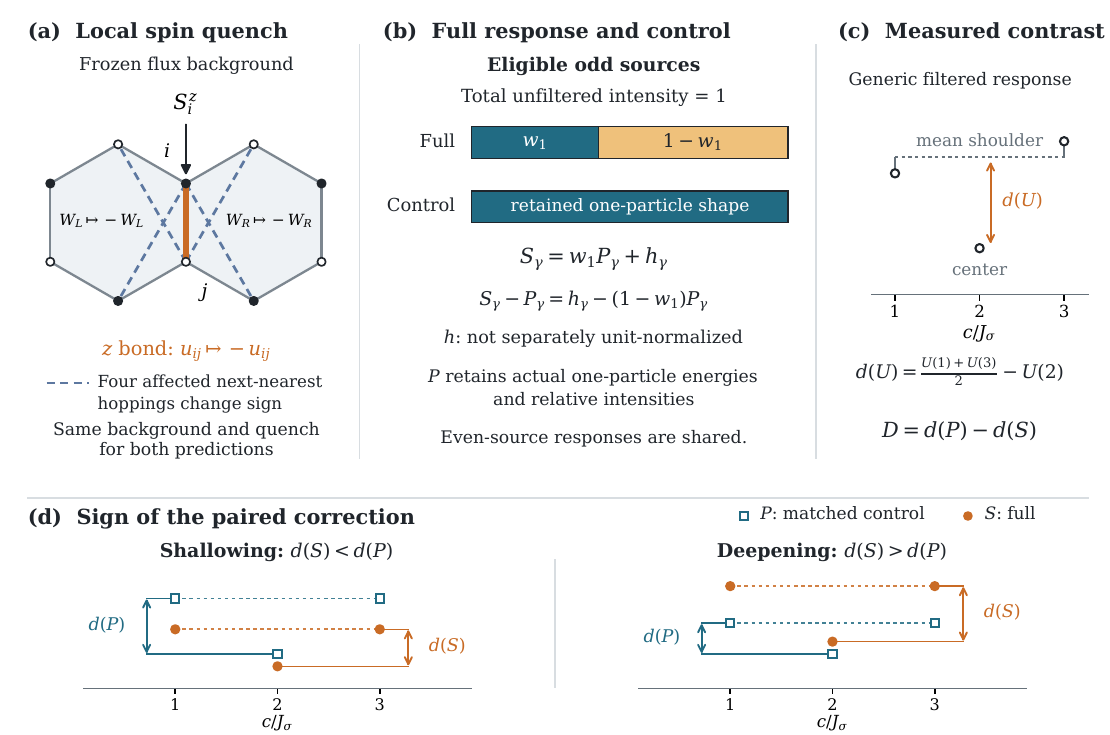}
\caption{\QuenchSchematicCaption}\label{fig:schematic}
\end{figure*}

Zhu \emph{et al.} found a separation of spectral branches near twice the Pauli exchange energy in their momentum-resolved disordered Majorana spectra and same-sublattice spin correlations~\cite{Zhu2025}. Their spin--Majorana comparison uses slow-flux and matter-overlap approximations. The branch separation motivates our choice of center frequency. We denote the Pauli exchange energy by $\Js=J/4$ for the spin convention $S^\alpha=\sigma^\alpha/2$ and coupling $J S_i^\alpha S_j^\alpha$. For the local spin response we choose three equally spaced frequencies, $\Js$, $2\Js$, and $3\Js$. At each frequency a filter gives the most weight to transitions nearest in energy. We use a narrow and a broad filter to compare the contrast at two energy resolutions. The flanking frequencies stay fixed for every spectrum, whether or not a visible trough occurs at the center. Here the local spin-transition amplitudes are calculated in the post-flip matter eigenbasis.

The comparison keeps the spin insertion, flux background, transfer-energy origin, and total intensity the same in both responses. Physical parity selects backgrounds with odd or even final quasiparticle number~\cite{Pedrocchi2011,Zschocke2015}. We call the state prepared by the spin insertion a \emph{source}. The one-particle replacement applies to odd sources with nonzero one-particle weight; on even sources both predictions retain the same full response. On each torus the eligible backgrounds, weighted by their relative disorder probabilities, define the conditional population. We take the depth difference on each of these sources before forming the conditional median. This median gives the typical depth correction at each size. At both energy resolutions, restoring multiparticle weight shallows the depression on 24 sites and deepens it on both 1,536-site shapes and on the 3,456- and 6,144-site tori [Fig.~\ref{fig:reversal}(a)]. Figure~\ref{fig:absolute} shows how the \emph{mean} intensities at the center and flanks change on the 24-site and 6,144-site tori.

Local-quench and few-particle descriptions, including sum-rule-normalized adiabatic comparisons, are established tools~\cite{Knolle2014,Knolle2015}. Arbitrary-flux quench correlations~\cite{Udagawa2018}, finite-temperature gauge sampling~\cite{Yoshitake2017}, field-driven spin and dimer spectra~\cite{Wang2025}, and adiabatic spin-response approximations in a Kitaev ladder~\cite{Feng2026} give related settings. For each eligible odd source, we compare the intensity-matched one-particle response with the full response, which includes transitions to final states with one, three, five, or any higher odd number of quasiparticles. The one-particle transition energies remain exact for that background, and a common rescaling preserves their relative spin intensities. This intensity-matched approximation is distinct from replacing the initial matter state by an adiabatically adjusted state or substituting a bare Majorana spectral function for a spin response. The disorder prescription stays fixed across sizes, so this series tests the finite-torus contrast error and its population dependence, rather than how that error changes as disorder is increased.

\section{Methods: local quench and intensity matching}\label{sec:model}

\subsection{Frozen Hamiltonian and spin-created source}

Let $u_{ij}=\pm1$ be the stored bond sign oriented from sublattice $A$ to $B$. The fixed-background matter Hamiltonian is
\begin{equation}
 H_u=i\Js\sum_{\langle ij\rangle_{A\to B}}u_{ij}c_i c_j
 + i\lambda\sum_{(i j k)_{\circlearrowleft}}
                 u_{ij}u_{jk}c_i c_k .
 \label{eq:hfrozen}
\end{equation}

Here $\{c_i,c_j\}=2\delta_{ij}$. The second sum follows consecutive edges counterclockwise around each hexagon, using the stored signs irrespective of traversal and including each term once. Equivalently, $H_u=(i/4)c^T A_u c$. An oriented bond $i\to j$ contributes $2\Js u_{ij}$ to $(A_u)_{ij}$, and an oriented path $i\to j\to k$ contributes $2\lambda u_{ij}u_{jk}$ to $(A_u)_{ik}$. The reverse entries have the opposite signs. We use $\lambda/\Js=1/4$ throughout.

Each stored link is negative independently with probability $p\simeq0.1965188845$, giving mean plaquette flux $(1-2p)^6\simeq0.05$ and allowing both Wilson-loop signs. Neighboring fluxes are correlated through their shared edge. This fixed distribution and coupling follow the disordered example of Ref.~\cite{Zhu2025}; its Pauli exchange $J$ is our $\Js$. They do not specify the field-dependent flux populations of the interacting magnet. The exact probability convention and lattice geometry are in Supplemental Material, Sec.~S1.

Diagonalizing $H_u$ gives independent fermion modes and a matter vacuum in which all are empty. The Majorana representation enlarges the spin Hilbert space. In a fixed link background, the physical spin constraints select the allowed matter-fermion parity~\cite{Pedrocchi2011,Zschocke2015}. If the vacuum has that parity, it is the initial state. Otherwise one occupation changes the parity, and filling a lowest-energy mode gives the lowest allowed energy. When the allowed ground space has an established degeneracy $g$, we prepare its normalized projector by giving each member of an orthonormal basis probability $1/g$.

A local $S^z_0$ insertion on sublattice $A$ applies $c_0$ with spin factor $1/2$ and changes $u$ to $u'$. It reverses the selected $z$ link, toggles the two adjacent fluxes, and reverses four dependent next-nearest-neighbor coefficients in Eq.~(\ref{eq:hfrozen}). For a pure initial ground state $|g_u\rangle$ of energy $E_{u,g}$, the onsite correlator in units $\hbar=1$ is
\[
 C^{zz}_{00}(t)=\tfrac14 e^{iE_{u,g}t}
 \langle g_u|c_0 e^{-iH_{u'}t}c_0|g_u\rangle .
\]
The first insertion prepares $c_0|g_u\rangle$, which propagates with $H_{u'}$ while the changed link signs remain fixed. The second insertion restores the original background. The phase comes from the initial eigenstate. Both spectra use this same intermediate evolution [Fig.~\ref{fig:schematic}(a)].

The initial state belongs to $H_u$, so applying a single $c_0$ can populate several quasiparticle-number sectors of $H_{u'}$. Label a final eigenstate $|I,u'\rangle$ by the set $I$ of occupied fermionic normal modes of $H_{u'}$. Their creation operators are linear combinations of the site Majoranas, and occupations are counted from the final vacuum. Three occupied modes, for example, specify one three-quasiparticle state. Inserting these eigenstates into the correlator gives the matter spectral measure
\begin{equation}
 d\mu(\omega)=\sum_I |\langle I,u'|c_0|g_u\rangle|^2
                   \delta(\omega-E_{u',I}+E_{u,g})\,d\omega .
 \label{eq:mu}
\end{equation}

Each squared overlap gives the intensity of a transition at its signed energy transfer $\omega=E_{u',I}-E_{u,g}$. Distinct final states at the same energy contribute additively. Since $c_0^2=1$, the measure has unit mass and the physical local $S^z$ intensity has total mass $1/4$. The parity of $c_0|g_u\rangle$ relative to the final vacuum selects either the odd or the even occupation numbers.

On a fixed flux background with a $g$-fold degenerate ground space, we assign equal probability to the $g$ states of an orthonormal ground-state basis. The intensity of a transition to a given final state is therefore the average of the squared spin-transition amplitudes from those initial states. This average depends only on the projector onto the ground space and is the same in any basis of that space. Equally probable initial states can nevertheless contribute different amounts of one-particle intensity. For example, if two such states place 80\% and 20\% of their respective intensities in one-particle lines, the first supplies four times as much one-particle weight as the second. We preserve that ratio by averaging the line intensities first, then rescaling the combined one-particle spectrum to the full integrated intensity. Rescaling each spectrum separately before averaging would give the two one-particle line shapes equal weight.

The transfer energy includes the change in raw-vacuum energy, the cost of any occupation required by the initial parity, and the final quasiparticle energies. We add the same constant to every frozen sector, so it cancels from each transfer. The mode energies and allowed parities still depend on the background, giving different physical ground-state energies. The mixture is stationary because each background retains its assigned probability and is prepared in its own ground space. A spin flip can connect an occupied sector ground state to a lower-energy state in another background, even to an excitation above that background's ground state. Such a transition releases energy to the probe and contributes at $\omega<0$ before broadening. We retain it at its signed transfer in every filter.

\begin{figure*}[t]
\includegraphics[width=.96\textwidth]{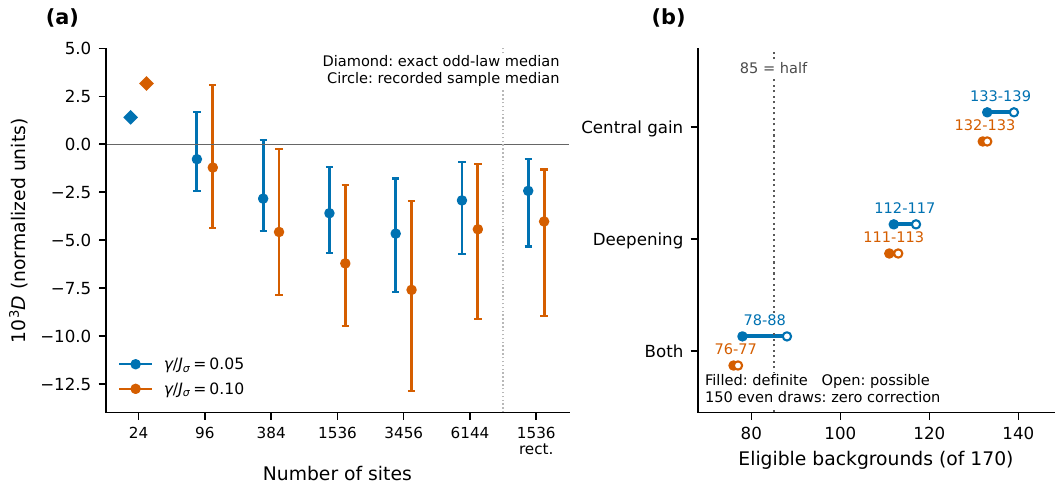}
\caption{\ReversalCaption}\label{fig:reversal}
\end{figure*}

\subsection{The approximation and the contrast measurement}

The full response $S$ includes every allowed quasiparticle sector. An odd source can populate one-, three-, five-, and higher odd-quasiparticle states. Parity alone does not ensure a nonzero one-particle transition amplitude. On an eligible odd source, let $w_1>0$ be the fraction of the unfiltered intensity in the one-particle sector. The intensity-matched approximation $P$ keeps that sector's actual transition energies and relative spectral weights, then multiplies the weights by $1/w_1$ so that its total intensity equals that of $S$. These weights are squared transition amplitudes, not the amplitudes themselves. The retained physical one-particle contribution is thus $w_1P$. We denote the combined three-and-higher-particle contribution by $h$; it carries total weight $1-w_1$, rather than being independently normalized. This construction is illustrated in Fig.~\ref{fig:schematic}(b).

To make each spectral reading at a finite energy resolution, we use the peak-height-one Lorentzian
\begin{equation}
 F_\gamma(\omega-c)=\frac{\gamma^2}{(\omega-c)^2+\gamma^2}.
 \label{eq:filter}
\end{equation}

Its half-width at half maximum, $\gamma$, is a chosen energy resolution, not an inferred lifetime. We use $\gamma/\Js=0.05$ and $0.10$. Writing $d\mu_1$ and $d\mu_h$ for the one-particle and combined higher-sector measures gives
\begin{align}
 S_\gamma(c)&=\int F_\gamma(\omega-c)\,d\mu(\omega), \nonumber\\
 P_\gamma(c)&=\frac{1}{w_1}\int F_\gamma(\omega-c)\,d\mu_1(\omega), \nonumber\\
 h_\gamma(c)&=\int F_\gamma(\omega-c)\,d\mu_h(\omega).
 \label{eq:filterdefs}
\end{align}

These source-normalized filtered intensities satisfy
\[
 S_\gamma(c)=w_1P_\gamma(c)+h_\gamma(c).
\]
An even source permits only even occupations and has no one-particle line to rescale. We use its full response in both predictions, giving zero paired correction even when its spin intensity is appreciable. Absolute spin intensities are $\absI_S=S/4$ and $\absI_P=P/4$; a unit-area spectral density would require an additional division by $\pi\gamma$.

For any filtered response $U$, define its depth relative to the average of the two flanks and the paired correction by
\begin{align}
 d(U)&=\tfrac12[U(1)+U(3)]-U(2),\nonumber\\
 D_\gamma&=d(P_\gamma)-d(S_\gamma).
 \label{eq:d}
\end{align}
Frequency arguments here and below are in units of $\Js$. The mean of the equally spaced flank readings is the value at the center of a straight line joining them. Subtracting the measured center cancels any constant level or linear slope. A positive depth means the center lies below this reference, even if the three readings increase monotonically. A positive $D$ means that restoring the higher sectors makes the contrast shallower, and a negative $D$ means deeper [Fig.~\ref{fig:schematic}(c,d)]. The additional readings at $1.8\Js$ and $2.2\Js$ check nearby central intensities with the two flanking frequencies held fixed.

The central change and the depth change follow from the same redistribution:
\begin{align}
 X_\gamma(c)&=S_\gamma(c)-P_\gamma(c)
       =h_\gamma(c)-(1-w_1)P_\gamma(c),\label{eq:x}\\
 D_\gamma&=X_\gamma(2)-\tfrac12[X_\gamma(1)+X_\gamma(3)].\nonumber
\end{align}
Restoring $h$ also removes the excess weight assigned to the one-particle lines by intensity matching. The central intensity can increase and the depression still deepen if the two flanks gain more intensity on average. At fixed $\gamma$, every line contributes the same Lorentzian area per unit weight as $c$ runs over the frequency axis. Equal total weight therefore gives $\int X_\gamma(c)\,dc=0$. A gain in one frequency range is balanced by a loss elsewhere, possibly outside the three measured bands. The sum rule thus permits either sign for the central change $X_\gamma(2)$ and for the depth correction $D_\gamma$. Accessible few-particle energy sums illustrate spectral support~\cite{ZschockeThesis}, but the coherent spin amplitudes determine its weight; Sec.~S1 gives their Gaussian construction and interference terms.

\begin{figure*}[t]
\includegraphics[width=.93\textwidth]{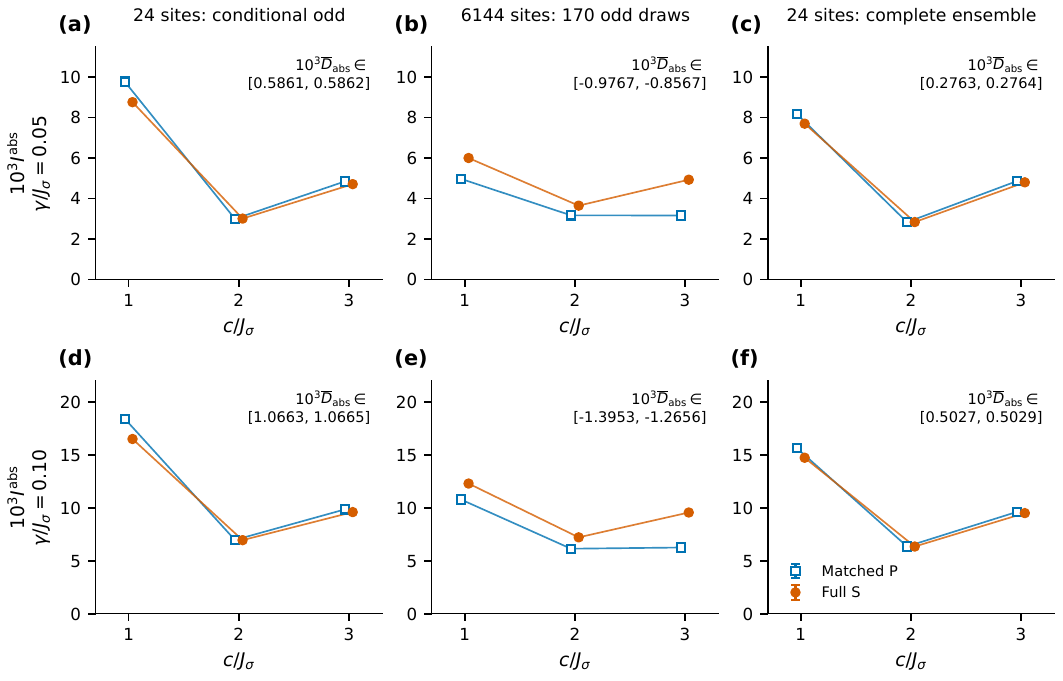}
\caption{\AbsoluteCaption}\label{fig:absolute}
\end{figure*}

\subsection{Finite tori and the paired statistic}

The reference torus has $N=24$ sites and translation columns $(2,2)$ and $(-2,4)$ in the primitive-cell basis. Scaling both columns by $n$ gives the homothetic sequence $N=24n^2$, through $96,384,1536,3456,6144$. A $24\times32$-cell rectangle supplies a second 1,536-site shape. At 24 sites we enumerate all 8,192 gauge orbits with their exact independent-link probabilities. For each larger size and shape we draw 320 independent assignments of the link signs under the same disorder law. Each assignment gives one frozen background. We reuse these draws for both spectra and both filter widths on that geometry. The 320 include even and odd sources. A draw number identifies a background only within its own geometry. Draws with the same number on different tori were generated independently.

We pair the two spectra on each eligible odd background $u$ and calculate $D_u=d(P_u)-d(S_u)$. On 24 sites we take the median of these corrections using the exact orbit probabilities conditioned on eligibility. On each larger torus the eligible draws sample its conditional population. We compare the resulting median corrections across geometries. Taking the difference first retains the relation between the two spectra on the same source, which is lost on taking their medians separately. The sampled median enclosures have joint coverage at least 95\% across the 48 tested combinations of geometry, filter width, and observable, under the stated disorder and numerical assumptions. The exact small-torus summaries have numerical but no disorder-sampling uncertainty.

For a fixed sample, we bound the mean dip-depth correction by averaging the lower and upper numerical bounds separately with the same background weights. On the 96-site torus with the narrow filter, the mean for the 164 eligible odd backgrounds lies in $[-0.686,-0.333]$, in units of $10^{-3}$ of the unit-normalized total intensity. Even the upper bound is negative, so the full response has a deeper mean dip than the matched one-particle response for this sample. A fresh sample from the same disorder distribution could contain a different mix of backgrounds and give a different mean. To estimate the mean over all eligible odd backgrounds on this torus, we resample the observed eligible backgrounds with replacement and propagate their numerical bounds. The approximate 95\% bootstrap interval for this geometry and filter width is $[-1.882,+0.901]$ in the same units (Table~S4). It spans zero, so both average deepening and average shallowing remain compatible with the data for the 96-site disorder population. Sections~S2 and S3 give the ranks, counts, numerical allowances, and full results.

\subsection{Tool use and verification}

OpenAI ChatGPT Astra and Codex configured with Astra assisted with theoretical and numerical analysis, coding, figures, and manuscript preparation under the author's direction. Scientific verification includes the numerical tests described in Supplemental Material, Sec.~S2, and independently implemented reductions of supplied numerical bounds and probability weights in Sec.~S6.

\section{The correction reverses across finite tori}\label{sec:reversal}

\subsection{Paired depth changes}

Each even source has exactly zero depth correction because both predictions use its full response. On 24 sites even sources carry 52.85\% of the disorder probability, enough to fix the exact all-source median at zero. On 6,144 sites 150 of the 320 draws are even. At each filter half-width fewer than 160 corrections are negative and fewer than 160 are positive, so the two middle entries in sorted order, 160 and 161, are both zero (Sec.~S2).

Including even sources gives a median dip-depth correction of zero on 24 sites and in the 6,144-site sample. Figure~\ref{fig:reversal}(a) shows what happens when we restrict the comparison to eligible odd sources. A background qualifies if the spin-created state has odd parity relative to the post-flip vacuum and carries nonzero one-particle weight. The vacuum and its one-particle modes depend on the flux background and the torus geometry, so the same rule selects a different population on each torus.

For each qualifying background, we subtract the full dip depth from the depth predicted by the intensity-matched one-particle approximation. We then take the median of those differences separately on each torus. At both filter widths, the median is positive on 24 sites, meaning that restoring multiparticle weight typically makes the dip shallower. On both 1,536-site shapes and on the 3,456- and 6,144-site tori, the median is negative at both filter widths, meaning that restoring multiparticle weight typically makes the dip deeper.

At 96 sites neither filter resolves the sign of the median depth correction, and at 384 sites only the broader filter resolves deepening. Table~S2 gives every median enclosure, including the wider numerical rows.

On 6,144 sites the median intensity change is positive at $1.8\Js$, $2\Js$, and $2.2\Js$ for both filter half-widths. The intensity at the center can rise while the depression deepens if the mean flank intensity rises more. These two effects need not occur together on every background. With the broader filter, fewer than half the realized odd backgrounds exhibit both signs, even though each marginal median has the stated sign [Fig.~\ref{fig:reversal}(b)].

\subsection{Mean spectra on the 24- and 6,144-site tori}

Figure~\ref{fig:absolute} shows how the three mean intensities produce the reversal in the odd-source mean. At each frequency the figure averages the intensities over the same backgrounds. Since the depth is linear in the three intensities, the depth of each mean spectrum equals the mean of its depths. As in Fig.~\ref{fig:schematic}(d), the depth is the vertical distance from the center to the line joining the two flanks. The following intensities and depths are in units of $10^{-3}$ absolute spin intensity. With the narrow filter on 24 sites both flanks lose intensity while the center barely changes, giving depths of about $4.31$ for the matched approximation and $3.73$ for the full response [Fig.~\ref{fig:absolute}(a)]. On 6,144 sites both flanks gain more intensity than the center. Their mean rises from $4.07$ to $5.47$ while the center rises from $3.17$ to $3.65$. The dip thus deepens from about $0.90$ to $1.82$ even as the central intensity rises [Fig.~\ref{fig:absolute}(b)]. The mean depth corrections on 24 and 6,144 sites are approximately $+0.586$ and $-0.917$ with the narrow filter, and $+1.066$ and $-1.330$ with the broader filter. Means and paired medians can differ even in sign. On 24 sites, for example, the narrow-filter central mean increases while its paired median decreases.

The mean fraction of intensity in higher-particle transitions grows from $0.12105$ in the exact 24-site odd-source distribution to $0.49080$ among the 170 eligible 6,144-site draws. Where that added intensity falls accounts for most of the change in the mean depth correction. With the narrow filter the higher sectors add slightly more intensity at the center than at the mean of the flanks on 24 sites. On 6,144 sites they add more at the flanks. Restoring these transitions also removes the excess intensity assigned to the one-particle lines to match the total. Of the change in mean correction from 24 to 6,144 sites, the added multiparticle intensity contributes $-1.547$ and undoing the one-particle rescaling contributes $+0.044$, for a net $-1.503$ in units of $10^{-3}$ absolute spin intensity (Table~S5).

To express the mean error relative to the full mean contrast, define
\begin{align}
 F_{N,\gamma}&=d(\langle S_\gamma\rangle_o),\nonumber\\
 M_{N,\gamma}&=\langle D_\gamma\rangle_o,\qquad
 R_{N,\gamma}=M_{N,\gamma}/F_{N,\gamma}.
 \label{eq:relativecontrast}
\end{align}
Both averages use the same eligible odd sources and their weights. On the 24- and 6,144-site tori, at Lorentzian half-widths $0.05\Js$ and $0.10\Js$, the full mean intensity at $2\Js$ lies below the average at $\Js$ and $3\Js$. This gives a positive reference depth $F$ in all four cases. Dividing the mean depth correction by the full mean depth gives a one-particle overestimate of approximately 16--18\% on 24 sites and an underestimate of 33--56\% in the 6,144-site sample across the two filter widths (Table~S7). For the fixed cohort of 170 eligible odd backgrounds on the 6,144-site torus, the numerical bounds give an underestimate of 45.6--55.7\% with the narrow filter and 33.4--38.3\% with the broader filter. A fresh disorder sample could give a different percentage.

The curves in Fig.~\ref{fig:absolute} and the percentages in Table~S7 use one-particle spectra rescaled to each source's full intensity before averaging. Table~S6 averages the unscaled one-particle spectra over the same sources and then rescales the average once. Sources with more original one-particle weight thus contribute more to the pooled line shape. Both prescriptions give a mean overestimate on 24 sites and a mean underestimate on 6,144 sites at both filter half-widths.

To test how the result depends on the mix of sources, we ask whether the large-torus depression would still deepen if one-particle-dominated sources were as common as on 24 sites. We group the 170 eligible 6,144-site draws into four equal-width intervals of $w_1$, the fraction of intensity in one-particle transitions before rescaling. Each group receives the probability of its 24-site counterpart, while each source keeps its calculated depth correction. For example, the 26 draws with $w_1\geq0.75$ together receive 90.64\% of the weight. At the narrow filter half-width the reweighted mean and paired median each allow either sign. At the broader half-width the mean becomes positive, giving shallowing on average (Table~S5). Reweighting by the number of negative adjacent plaquettes instead preserves both mean and paired-median deepening with the narrow filter. The primary result uses the original disorder distribution conditioned on odd eligibility. Reweighting backgrounds by $w_1$, which is itself an outcome of the quench, demonstrates composition sensitivity without isolating its microscopic cause.

A transition away from $2\Js$ can contribute to the central intensity through the tail of the Lorentzian. Doubling the filter half-width from $0.05\Js$ to $0.10\Js$ leaves the contribution of a line at the center unchanged because the peak height is fixed. A distant line contributes nearly four times as much. Comparing the two readings therefore bounds the fraction of the narrow-filter central intensity that comes from transitions within $0.10\Js$ of the center, separately for the one-particle and higher-particle sectors. Background 003 is one of the 170 eligible odd sources on the 6,144-site torus. On this background, transitions within the window account for at least 25\% of the higher-particle reading and at most 4.7\% of the one-particle reading. The minimum higher-particle fraction exceeds the maximum one-particle fraction on 37 of the 170 eligible 6,144-site backgrounds. None forces the reverse ordering, and 133 remain unresolved (Sec.~S4 and Fig.~S1). This comparison locates weight contributing to the central reading. The depth also depends on the flanks. A locally flat density and a single suitably detuned line can both give a twofold increase, so the two readings alone cannot establish a continuum.

\begin{figure}[tb]
\includegraphics[width=\columnwidth]{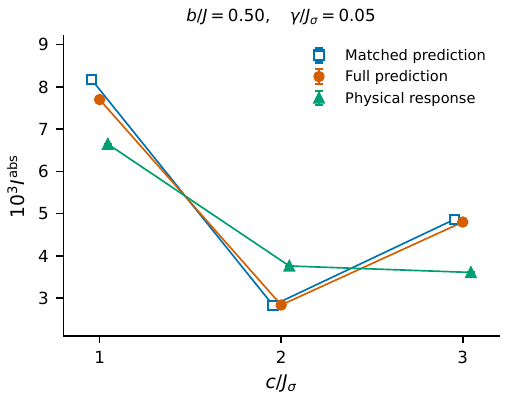}
\caption{\BenchmarkCaption}\label{fig:benchmark}
\end{figure}

\section{A secondary benchmark in the interacting magnet}\label{sec:field}

The comparison within the frozen ensemble isolates the effect of restoring multiparticle transitions. An independent calculation on the same 24-site torus tests whether restoring them also improves the predicted local spin intensities of the interacting magnet in a $[111]$ field. Its Hamiltonian is
\begin{equation}
 H=J\sum_{\langle ij\rangle_\alpha}S_i^\alpha S_j^\alpha
   -\frac{b}{\sqrt3}\sum_i (S_i^x+S_i^y+S_i^z),
 \quad S_i^\alpha=\sigma_i^\alpha/2 .
 \label{eq:magnet}
\end{equation}

Here $J=1/2$ and $\Js=1/8$. We compare both frozen predictions with the local $S_0^z$ response of the interacting magnet at $b/J=0.50$. The spin sum rule fixes the integrated intensity at $1/4$ and thus sets the one-particle normalization independently of the interacting data. All three spectra use signed transfer $\omega=E_f-E_i$ and include elastic weight. Equal additive offsets in the frozen sectors preserve their calculated energy differences. We fixed the frozen hopping $\lambda/\Js=1/4$, the background probabilities, the three probe frequencies, and the filter half-width before evaluating the interacting intensities. The predictions therefore enter the comparison on an absolute scale, without adjusting their amplitudes or frequencies.

Figure~\ref{fig:benchmark} compares the three intensities at $c/\Js=1,2,3$ and $\gamma/\Js=0.05$. Restoring the higher sectors reduces the prespecified unweighted sum of squared errors by 28.8--29.7\%, chiefly through the left flank. Both frozen predictions still place the right flank above the center, whereas the interacting response places it below. In the full frozen ensemble, transitions with $E_f<E_i$ carry approximately 13.72\% of the total local spin intensity before broadening. The initial state has the lowest allowed energy in its own sector, but the spin flip can reach a lower state in the new sector. The negative transfers are present at the line energies, before Lorentzian tails enter the spectrum. For a true ground state every final energy satisfies $E_f\ge E_0$. Section~S5 gives the intensity errors at each probe frequency, compares the mean energy cost of a spin flip for each initial flux sign of an adjacent plaquette, and presents the numerical evidence for identifying the interacting reference state as the ground state.

We next ask whether changing only the probabilities of the frozen backgrounds can reproduce selected observables of the interacting 24-spin state while eliminating downward transfers for this probe. We use the same numerical state at $b/J=0.50$ as in Fig.~\ref{fig:benchmark}. Before the spin flip, the two hexagons sharing the flipped bond have four possible ordered pairs of plaquette flux signs, $(++,+-,-+,--)$. Their probabilities measure the local flux correlations in that state. For the spectral measurement, we instead resolve the state immediately after the spin flip by the flux sign $+1$ or $-1$ on just one of those hexagons.

Let $A_+$ and $A_-$ be the amplitudes from the two resulting flux components to the same final energy eigenstate. The $[111]$ field mixes flux sectors, so both components can reach that state. A direct energy measurement gives intensity $|A_++A_-|^2$. If one first measures the plaquette flux and discards the outcome, the two components retain their original probabilities and give the dephased intensity $|A_+|^2+|A_-|^2$. The difference, $2\,\mathrm{Re}(A_+^*A_-)$, depends on their relative phase. Equal in-phase amplitudes $a$ give $4|a|^2$ coherently and $2|a|^2$ after dephasing. Opposite amplitudes cancel coherently and still give $2|a|^2$ after dephasing. Summing the dephased intensities over final states with filters centered at $\Js$, $2\Js$, and $3\Js$ gives the three spectral targets, including transitions at zero energy transfer.

Each of the 8,192 frozen gauge orbits groups equivalent bond-sign configurations into one physical background. A background has a definite plaquette flux, so dephasing leaves its intensity unchanged. After the interacting targets were known, we varied only the orbit probabilities, keeping the preparations, transition energies, and spin matrix elements fixed. Section~S5 verifies that one set of seven nonnegative rational weights sums exactly to one and places all four pair populations and all three dephased intensities within their respective numerical bounds. Each occupied orbit has zero downward weight for this probe. The fit establishes that static populations can reproduce these selected data. It cannot determine the interacting populations independently or produce the relative phase between the two flux components.

Section~S5 also compares two fields in the interacting 24-spin magnet. As $b/J$ rises from $0.50$ to $0.55$, the minimum on the sampled frequency grid moves up in energy by less than the bare Zeeman shift. At the fixed center $2\Js$ and half-width $0.05\Js$, the difference between the coherent and dephased readings after elastic subtraction gives the interference contribution to the inelastic spin intensity. This contribution is positive and grows with field, while bare spin precession predicts a decrease. Kaib, Winter, and Valent\'{\i} resolved the plaquette-flux fluctuations themselves in momentum and energy in a $[111]$ field~\cite{Kaib2019}. The dephasing comparison measures how coherence between the two flux signs of one plaquette contributes to the local spin intensity at a specified energy.

\section{Discussion and conclusion}\label{sec:conclusion}

Matching the integrated spin intensity does not fix how a one-particle spectrum distributes that intensity around a local depression. In the frozen, flux-disordered Kitaev honeycomb model studied here, restoring the higher-particle sectors changes the conditional median correction from shallowing on 24 sites to deepening on both 1,536-site shapes and the 3,456- and 6,144-site tori, at both energy resolutions. The same local quench, retained one-particle energies, relative spin weights, and total intensity enter each paired comparison. The 24-site torus alone would thus predict the wrong direction of the typical depth correction on all four larger tori. The population of eligible sources also matters. We sort the 6,144-site sources into four groups by their one-particle intensity fraction and assign each group its 24-site probability. Each source keeps its own depth correction, yet the bounds on the reweighted median now allow either shallowing or deepening with the narrow filter.

The mean depths on the 24-site and 6,144-site tori give the size of the error. When we normalize the one-particle spectrum on each eligible odd source before averaging, the one-particle approximation overestimates the full mean depth by about 16--18\% on 24 sites and underestimates it by 33--56\% in the 6,144-site sample, relative to the full mean depth in each case. Averaging the original one-particle spectra first and then rescaling the pooled spectrum to the total intensity of the full mean spectrum gives greater weight to sources with a larger one-particle fraction. At both filter widths this order of normalization also gives too deep a mean depression on 24 sites and too shallow a mean depression in the 6,144-site sample.

A future extension toward scattering observables is a quench-aware comparison of full and one-particle momentum-resolved spin structure factors for extensive flux configurations, using center-of-mass averaging for disordered backgrounds, as illustrated for Majorana spectra in Ref.~\cite{FengTrivedi2026}, or a periodic flux crystal. Such a comparison would show whether the redistribution of spin intensity near $2\Js$ is spread across the Brillouin zone or concentrated near particular wave vectors.

\vfill
\section*{Data availability}
Numerical bounds, exact small-system probability weights, reduction scripts, and supporting records identified in Supplemental Material, Sec.~S6, are publicly available in the \href{https://github.com/davidronquillo/kitaev-spin-spectral-contrast}{\texttt{kitaev-spin-spectral-contrast}} companion repository. These materials are also supplied as a standalone numerical collection. Validation of the spectral calculations is described in Sec.~S2; the collection README specifies the supplied inputs, commands, and expected outputs.

\section*{Acknowledgments}
The author thanks Shi Feng, Yuan-Ming Lu, and Michael Peterson for their helpful comments and suggestions. The author directed the research and takes responsibility for the work. OpenAI ChatGPT and Codex assisted with literature review, theoretical and numerical analysis, coding, figures, and manuscript preparation.

\vfill
\interlinepenalty=10000
\bibliography{references}
\end{document}


{\large\bfseries Supplemental Material}\\[3pt]
{\Large\bfseries Finite-size reversal of a multiparticle correction to local spin spectra in a flux-disordered Kitaev honeycomb model}\\[6pt]
\begingroup
\renewcommand{\thefootnote}{\fnsymbol{footnote}}
{\large David C. Ronquillo\footnote{Correspondence: \href{mailto:david.ronquil@gmail.com}{david.ronquil@gmail.com}.}}
\endgroup
\par\smallskip
This supplement specifies the model, spin-created sources, energy reference, and probability conventions (Sec.~S1); numerical methods and inference (Sec.~S2); and the full finite-size evidence, including normalization and composition checks (Sec.~S3, Tables~S1--S7). Section~S4 gives the two-resolution support diagnostics and the prospective one-source test. Section~S5 preserves the secondary interacting benchmark, seven-orbit fit, and field-displacement/dephasing evidence. Section~S6 identifies the accompanying numerical collection and reproduction capabilities. Figures~S1 and S2 give the width/window and field-displacement diagnostics, respectively.

\secS{1}{Model, source, and energy reference}

Use primitive vectors $\mathbf a_1=(\sqrt3/2,3/2)$ and $\mathbf a_2=(\sqrt3,0)$, with basis offsets $\delta_A=(0,0)$ and $\delta_B=(0,1)$. In cell coordinates an $A_R$ site has neighbors $B_{R+d_\alpha}$ with $d_x=(0,0)$, $d_y=(-1,1)$, and $d_z=(-1,0)$. A counterclockwise hexagon, in cyclic order, is
\[
 A_R,\ B_R,\ A_{R+(1,-1)},\ B_{R+(0,-1)},\ A_{R+(0,-1)},\ B_{R+(-1,0)}.
\]
The product of the six stored signs gives its flux. Products of two consecutive signs along this oriented hexagon give the next-nearest-neighbor coefficients in Eq.~(1) of the article. There is no extra Hermitian-conjugate term. The complete ensemble includes the noncontractible sectors induced by the original link signs. The exactly implemented negative-link probability is
\[
 p=\frac{66871911161826185187153288884631757060}{2^{128}},
\]
the $128$-bit discretization of $(1-20^{-1/6})/2$. Each fresh sample ID, geometry, cell, and bond specifies a different counter input. Statistical coverage assumes these reproducible draws represent independent Bernoulli links. The accompanying source records specify the graph labeling, parity ordering, and sampling convention; the scalar reductions use the recorded geometry, source identity, and eligibility.

The fixed-gauge representation $\sigma_j^\alpha=i b_j^\alpha c_j$, with local constraint $D_j=b_j^x b_j^y b_j^z c_j=+1$, fixes the allowed global matter parity after reordering the Clifford operators on the labeled graph. The ordering sign must be evaluated for the actual torus. For a nonsingular matter block, the positive normal modes of $iA_u$ specify the raw vacuum and its parity. When the raw vacuum is disallowed the lowest allowed space has one lowest-energy matter occupation. If the lowest allowed energy has a verified degeneracy $g$, the preparation is $\rho_u=g^{-1}\sum_{a=1}^g |a,u\rangle\langle a,u|$, where the $|a,u\rangle$ form any orthonormal basis of that ground space. Each basis state contributes its full spin-transition measure with weight $1/g$. The local $z$ insertion reverses $u_z(0,0)$, four dependent next-neighbor hoppings, and the two adjacent faces. Both model predictions use the same quenched matrix and the same transformed source.

For an initial occupation $n_i=0$ or $1$ and positive mode energies $\epsilon_m^{i,f}$, write
\begin{align}
 \Omega_I&=\sum_{a\in I}\epsilon_a^f,\\
 \Delta&=-\tfrac12\sum_m(\epsilon_m^f-\epsilon_m^i)-n_i\epsilon_{\min}^i,\qquad
 \omega_I=\Omega_I+\Delta .
\end{align}
The auxiliary $N=1152$ three-particle comparison used $\Omega_I$, measured above each final raw matter vacuum. The main size sequence and the three-band physical-model benchmark use the signed transfer $\omega_I$ from the initial allowed-state energy. A parity correction already included in that energy is applied once. Negative and zero transfers are retained. The additive energy offsets of the frozen-flux sectors are taken equal and were fixed before the physical comparison, without fitting them to that response. Equal sector energy offsets do not imply equal vacuum or ground-state energies: the background-dependent raw-vacuum energies and physical-parity corrections remain included. Changing the relative sector energy offsets would change the filtered spectra.

For a pure Gaussian preparation, $c_0$ transforms its Majorana covariance by the coordinate reflection $D_0=1-2e_0e_0^T$, giving $\Gamma_\chi=D_0\Gamma_iD_0$. In the final particle basis an odd Gaussian state has one blocked occupied mode and paired factors. For pair occupations $p_k$ its particle-number generating function is $z\prod_k(1-p_k+p_k z^2)$. This evaluates integrated sector masses without inferring energy-resolved weights from line counts. The actual $q=3$ intensities require coherent amplitudes, as in Eq.~(\ref{eq:wedge}) below. For a projector preparation let $\mu_{1,a}$ be the one-particle measure from $|a,u\rangle$ and $w_{1,a}$ its integrated mass. The matched measure is $(\sum_a\mu_{1,a})/(\sum_a w_{1,a})$ when the denominator is positive. The common factor $1/g$ cancels, so a component with zero one-particle weight contributes zero to both sums. The matched odd-source control retains the actual post-quench one-particle transfer energies, preserves their relative spectral weights, and divides that sector by its positive mass $w_1$ to restore the total intensity. Unlike an adiabatic spectral proxy that approximates overlaps or the spin vertex, this control gives the one-particle description the exact retained quench lines and their relative weights within the numerical allowances. The comparison isolates the spectral-shape correction produced by restoring higher sectors and undoing the one-particle rescaling within the stated frozen ensemble.

Here $\gamma$ is the Lorentzian filter half-width at half maximum, with settings $\gamma/\Js=0.05$ and $0.10$. It is a chosen energy-resolution parameter, not an inferred intrinsic linewidth or lifetime. For an eligible odd source, $S_\gamma(c)$ denotes the full filtered spin response in the source-normalized convention of the article, $P_\gamma(c)$ denotes the one-particle response rescaled to match the total local intensity, and $h_\gamma(c)$ denotes the combined contribution from the allowed higher sectors $q=3,5,\ldots$. Thus
\[
 S_\gamma(c)=w_1P_\gamma(c)+h_\gamma(c).
\]
The weighted contribution $w_1P_\gamma(c)$ retains its actual one-particle mass, while the control $P_\gamma(c)$ has unit unfiltered mass. The higher-sector measure has mass $1-w_1$ and is not separately normalized to unity. Multiplication by $1/4$ converts these filtered intensities to absolute single-component spin units. Even sources retain their full response in both predictions.

\textbf{Coherent particle-sector amplitudes.} The spin vertex fixes the relative intensities of the retained quench lines. In a regular odd Gaussian chart, write the spin-created state as a blocked creation operator with coefficients $\ell_a$ multiplying a paired exponential with antisymmetric matrix $Z_{ab}$. The one- and three-particle amplitudes have the form
\begin{align}
 M_a&={\cal C}\ell_a,\nonumber\\
 M_{abc}&={\cal C}(\ell_a Z_{bc}-\ell_b Z_{ac}+\ell_c Z_{ab}).
 \label{eq:wedge}
\end{align}
The three permutations describe the \emph{same} final occupation and interfere. Adding the densities of accessible energy sums cannot reproduce these amplitudes. Changing only five hopping coefficients is a local perturbation even in a large sample, and its effect on the local matrix element can stay finite.

\textbf{Blocked Gaussian construction.} The implemented chart uses the normal and anomalous contractions of the spin-created source in the retained final positive-energy frame $V$:
\[
 G_\chi=\tfrac12(I+i\Gamma_\chi),\qquad
 R=V^\dagger G_\chi V,\qquad K=V^\dagger G_\chi V^* .
\]
In exact canonical coordinates an eligible odd source has a unit eigenvector $\ell$ of $R$ with eigenvalue one. Set $P_\ell=\ell\ell^\dagger$ and $Q_\ell=I-P_\ell$. The implementation solves a deflated system, enforces ordinary-transpose antisymmetry, and removes the blocked component:
\[
 \widetilde Z=(I-R+P_\ell)^{-1}K,\qquad
 Z=Q_\ell\frac{\widetilde Z-\widetilde Z^{\mathsf T}}{2}Q_\ell^{\mathsf T},
 \qquad {\cal C}=\det(I+ZZ^\dagger)^{-1/4}.
\]
Thus the source is ${\cal C}(\sum_a\ell_a f_a^\dagger)
\exp[\tfrac12\sum_{ab}Z_{ab}f_a^\dagger f_b^\dagger]|0_f\rangle$.
The normalization is evaluated from the singular values $s_j$ of $Z$ through
$\log |{\cal C}|^2=-\tfrac12\sum_j\log(1+s_j^2)$; $w_1=|{\cal C}|^2$ for this normalized pure odd chart.
The numerical construction Hermitianizes $R$, antisymmetrizes $K$, and retains contraction, linear-solve, projection, normalization, and final-frame errors. It verifies the reconstructed normal and anomalous contractions. A nonisolated blocked direction, an unresolved unblocked inverse, an unresolved occupied/complementary spectral separation, or normalization underflow is retained as unresolved; no arbitrary mode or regularization replaces it. Orthogonality errors are bounded through a nearby real polar frame, without changing the retained chart's tuple amplitudes. The source, pure-chart reconstruction, and final-frame alignment contributions are combined before conversion of the covariance Frobenius allowance to a trace-distance allowance by division by $\sqrt8$. The separate one-body operator allowance includes mixing within the positive subspace; its $q$-particle effect is bounded by $q$ times that allowance, with the common transfer-offset uncertainty charged separately. At $N=24$, degenerate allowed initial spaces use all projector components with equal weights; their unnormalized measures are averaged before intensity matching. The large-source construction requires its admitted occupied space to be resolved. These conventions also prevent an unresolved degeneracy from being silently replaced by a singleton.

\textbf{Auxiliary sector and energy-axis comparisons.} The three-particle sector can carry a larger central fraction of its own total intensity than the one-particle sector. On a 1,152-site torus, we normalized each sector's filtered intensity at $2\Js$ by its total weight and formed the three-minus-one difference on each background. Across 12 odd sources, the median paired difference was $0.004499$, with conditional population-median interval $[0.0009496,0.0092995]$. Here energies were measured above each final raw matter vacuum, rather than as signed transfers from the initial state. A separate 1,800-site cohort retained positive central-intensity and negative paired depth-correction medians on both the final-vacuum excitation-energy axis and the signed-transfer axis. The following paragraph distinguishes the realized-sample excess from the population-median sector comparison.

\textbf{Sector-resolved comparison on 1,152 sites.} For the twelve eligible odd sources among twenty, the narrow-filter median full-minus-matched excess at $2\Js$ lay in $[0.0011447,0.0235831]$ even allowing every placement of the omitted higher-sector weight. This interval describes that realized sample. The separate three-minus-one statistic quoted above compares the filtered fractions after normalizing the $q=3$ and $q=1$ sectors separately to unit mass; its interval estimates the population median of their paired difference, conditional on odd eligibility. These auxiliary comparisons measure excitation energy above each final raw matter vacuum, unlike the signed transfers used for the main size sequence.

\par\medskip
\section*{S2.\quad Numerical evidence and inferential targets}
\addcontentsline{toc}{section}{S2. Numerical evidence and inferential targets}

\textbf{Finite-system sampling.} The reference uses translation columns $M=[(2,2),(-2,4)]$ and has $N=24$ spin or Majorana sites in the corresponding calculation. There are $2^{13}=8192$ gauge orbits including Wilson loops. The independent Bernoulli link probabilities specified in Sec.~S1 induce exact rational probabilities on these gauge orbits; those probabilities weight the exact average. Of the 8192 orbits, 3892 have an eligible odd source, with total probability $p_{\rm o}\simeq0.47149174$. The even-source probability $1-p_{\rm o}>1/2$ therefore places more than half of the unconditional probability exactly at zero correction for each $X(c)$ and for $D$. Their unconditional 24-site medians are identically zero. The final occupation enumeration uses all 2048 occupations in the allowed parity sector per 24-site matter block. The initial allowed spaces are singletons on 8176 orbits, doublets on twelve, and triplets on four.

Each independently drawn larger cohort has 320 assignments. In the order $N=96,384,1536,3456,6144$ on the tilted series and the $N=1536$ rectangle, the eligible odd counts are $164,156,175,177,170,135$. The remaining 943 sources are shared even cases. The numerical method and validation are summarized below. The positive filter ordinates are formed from the exact coherent $q=1,3$ sums within numerical allowances, importance estimates of higher sectors when needed ($q=5,7$), and bounds on unrepresented weight. They share a source, final eigensystem, and transfer offset within a background. The direct signed mean-shoulder-depth and filter-width-difference kernels retain correlations across ordinates. Numerical source and denominator uncertainty, energy registration, stochastic integration, and allowed placement of omitted mass all enter the final intervals. Independent full-source evaluations use a different Gaussian transform on specified check rows.

\textbf{Higher-sector importance integration.} For $q=2k+1=5,7$, a proposal draws one blocked-mode index and $k$ independent unordered mode pairs. The singleton and pair probabilities, $\pi_a$ and $\rho_{bc}$, approximate weights $|\ell_a|^2$ and $|Z_{bc}|^2$: each atom receives a positive integer mass, the masses sum to $2^{52}$, and the recorded integer probabilities enter the denominator exactly. Specifically, for $d$ atoms of weights $v_j$, the initial masses are $1+\lfloor(2^{52}-d)v_j/\sum_i v_i\rfloor$; the remaining tickets are assigned to the largest-weight atom. Collisions between selected indices contribute a zero observation and remain in the total proposal count.

Let $\mathcal M(I)$ be the $q!!$ signed singleton/pair matchings of a distinct ordered $q$-mode set $I$. If $a$ is the singleton and $(b,c)$ its matching pairs, the probability of that unordered occupation is
\[
 Q_q(I)=k!\sum_{m\in\mathcal M(I)}\pi_{a(m)}
                   \prod_{(b,c)\in m}\rho_{bc} .
\]
Its coherent weight is $W_I=|{\cal C}|^2
|\sum_m\operatorname{sgn}(m)\ell_{a(m)}\prod_{(b,c)\in m}Z_{bc}|^2$.
For a prespecified filter or signed kernel $H$, the observation is
$Y_H=W_I H(\omega_I)/Q_q(I)$. The same tuple supplies all 24 prespecified kernels, including both filter widths and the direct depth contrasts. With
$A_\ell=\max_a |\ell_a|^2/\pi_a$ and
$A_Z=\max_{b<c}|Z_{bc}|^2/\rho_{bc}$, the Cauchy bound is
\[
 0\le W_I/Q_q(I)\le B_q,
 \qquad B_q=|{\cal C}|^2\frac{q!!}{k!}A_\ell A_Z^k .
\]
The code enlarges $B_q$ and the observations for normalization and arithmetic error. If $H\in[h_-,h_+]$ with $0\in[h_-,h_+]$, its observation-range length is $R_H=B_q(h_+-h_-)$, using the enlarged cap. At cumulative counts $n=2^{14},\ldots,2^{25}$ it uses
\[
 r_n=\sqrt{\frac{2\widehat v_n^+\log(4/\alpha)}{n}}
       +\frac{7R_H\log(4/\alpha)}{3(n-1)}+\epsilon_{\rm arith},
 \qquad \alpha=\frac{0.01}{1920\,2\,12\,32}.
\]
Here $\widehat v_n^+$ bounds the unbiased sample variance including per-draw arithmetic, and $\epsilon_{\rm arith}$ covers the mean and reduction arithmetic. The 24 used slots leave eight unused; repeated looks refine the same source, not the disorder sample. Represented intervals are intersected with their sector-mass ranges and prior allocated looks. Source, energy, and common-offset allowances then give physical intervals. Any unrepresented mass is multiplied by the feasible kernel range, retaining both signs for signed contrasts.

\textbf{Recorded validation.} The following checks compare independent constructions on identical inputs. Discrepancies and allowances are in the source-normalized convention (total matter mass one); the spin intensities carry the separate factor $1/4$.
\begin{center}\small
\begin{tabularx}{\linewidth}{@{}p{.24\linewidth}X p{.29\linewidth}@{}}\toprule
Case & Comparison and recorded discrepancy & Recorded allowance or outcome\\\midrule
$N=24$, eight orbits & Occupation probabilities from explicit 4096-dimensional Fock kets versus Gaussian-source probabilities, including rank-two and rank-three initial projectors; largest component occupation-probability $L^1$ difference $4.19\times10^{-15}$. & Component $L^1$ allowances at least $8.79\times10^{-10}$; all compared boxes intersect. Largest numerical-enclosure half-width $4.53\times10^{-7}<10^{-6}$.\\[4pt]
Seven-mode complex source, $q=5,7$, three phase choices & Coherent matching amplitudes versus independent Fock creation; largest amplitude discrepancy $8.51\times10^{-17}$. & Amplitude tolerance $2\times10^{-12}$; 620 saved proposal, kernel, and stream checks passed.\\[4pt]
Ten-mode complex source, both filter widths & Full Fock characteristic transform versus independent Pfaffian/Wick transform, including positive and negative transfer shifts; largest recorded discrepancy $2.41\times10^{-15}$. & Allowance on that comparison $8.34\times10^{-11}$; all 406 saved transform, filter-bound, and shift checks passed.\\[4pt]
First eligible source on each of six larger geometries & Sector estimates versus independent-transform intervals after transport to the same represented source; no recorded incompatible comparison. & All eight direct-only targets met on three of six sources; all six combined checks met $2.5\times10^{-4}$. Wider direct-only checks retained.\\\bottomrule
\end{tabularx}
\end{center}
The eight small-system fixtures contain even sources, odd singletons, and odd degenerate preparations; their exact-zero-mode stratum was empty. A separate homogeneous-$N=24$ assembly control remained unresolved when occupied and complementary spectral intervals did not separate; it was not counted as a resolved response. The accompanying validation records give these case-specific comparisons and their numerical allowances.

\textbf{Protected medians.} On the fresh cohorts, 24 primary endpoints are $X(2)$ and $D$ at both filter widths on six geometries. The other 24 are $X(1.8)$ and $X(2.2)$. The disorder error allocations are $0.03$ and $0.01$ for the respective families. The two-sided single-endpoint levels are $\alpha=1/800$ and $1/2400$. For $m$ eligible draws, choose the largest $k$ satisfying
\begin{equation}
 2\sum_{j=0}^{k-1}\binom{m}{j}\,2^{-m}\le\alpha .
\end{equation}
For row boxes $[l_i,u_i]$, the population-median enclosure is $[l_{(k)},u_{(m-k+1)}]$, sorting lower and upper endpoints separately. At $N=6144$ the primary ranks are 64 and 107, and the secondary ranks 62 and 109. A further 0.01 global allowance covers fresh stochastic numerical enclosures across the permitted looks. The prespecified joint coverage floor is 95\% under the iid disorder and numerical assumptions. The reported discrete union-bound floor is 0.9614372. The order-statistic guarantee covers the fresh median family. The row-ratio summaries, simultaneous-event frequencies within the realized sample, and the window-fraction count have no additional population-coverage claim.

\textbf{Means and simultaneous signs.} Average the lower and upper physical row bounds with their original weights to enclose a fixed-cohort mean. The reported 50,000 whole-row percentile bootstrap resamples are approximate, pointwise population-mean diagnostics. For a pair of signs such as $X(2)>0$ and $D<0$, a row certainly qualifies if both entire boxes satisfy the inequalities. It can be counted as possible if both boxes allow them. On 6144 sites the narrow-width definite/possible counts are $133/139$ for $X(2)>0$, $112/117$ for $D<0$, and $78/88$ for their conjunction. With the broader filter they are $132/133$, $111/113$, and $76/77$, respectively. The 150 even draws have exactly zero correction. The odd-source sign counts complete the rank argument: with the narrower filter there are at most 117 negative and at most $170-112=58$ positive corrections; with the broader filter the corresponding upper bounds are 113 and $170-111=59$. Each sign therefore occurs fewer than 160 times in the full cohort. Both middle ranks, 160 and 161, are zero, even if additional odd corrections vanish. This establishes the all-background sample median at both filter widths. The count of 150 even draws alone would not establish it.

\textbf{Retained imprecision.} Of 977 fresh eligible rows, 936 met all eight integration-plus-tail goals, each a numerical-enclosure half-width of $0.00025$. The row counts are $164/164,154/156,171/175,164/177,151/170$ on the tilted series and $132/135$ on the rectangle. For example, the $N=6144$ broad-filter depth on ID233 has integration-plus-tail enclosure half-width $0.002351$ and the full numerical depth box $[-0.04693,-0.04221]$. It still resolves a negative sign. The first eligible direct-only check met all eight targets on $N=96,384,6144$. The two $N=1536$ checks and $N=3456$ each missed at least one direct-only target. All six combined checks passed on the common represented source, with no recorded method disagreement. None of the wider rows was removed.

The earlier independent cohort was refined deterministically on 189 historical rows without new draws or new Monte Carlo proposals. Its $N=1536$ tilted-cohort means then resolved positive $X(2)$ and negative $D$ at both widths. Four historical rows missed the joint two-filter goal. Three were tilted $N=1536$ IDs 062, 156, and 308; the fourth was rectangular $N=384$ ID185. Historical and fresh cohorts retain separate inference.

\par\medskip
\secS{3}{Absolute spectra, population medians, and finite-size stability}

\noindent\begin{minipage}{\linewidth}
\textbf{Table S1. Absolute model ordinates.} Entries are $10^3$ times the filtered local \emph{absolute spin intensity}, with the total unfiltered intensity equal to $1/4$. Here $\gamma$ is in units of $\Js$ and numerical center labels are $c/\Js$. The two $N=24$ rows per filter width are exact probability-weighted means over all backgrounds or conditional on odd eligibility. Larger-system rows are fixed-cohort odd-source means with numerical enclosures. “Rect.” marks the shape control. The displayed marginal boxes are rounded separately; the tighter depth annotations and their original-record reductions are discussed below.
\begin{center}
\resizebox{\linewidth}{!}{\begin{tabular}{@{}lccccccc@{}}\toprule
Sites / ensemble & $\gamma$ & $P_1/4$ & $S_1/4$ & $P_2/4$ & $S_2/4$ & $P_3/4$ & $S_3/4$\\\midrule
\AbsoluteRows
\bottomrule\end{tabular}}
\end{center}
\end{minipage}

The mean-shoulder depth $d(U)=[U(1)+U(3)]/2-U(2)$ is a linear functional. It does not require each shoulder individually to exceed the center. The neighboring central-intensity checks at 1.8 and $2.2\Js$ do not test robustness to changing the shoulders. Writing $o$ for odd eligibility and $p_{\rm o}=\Pr(o)$ gives $\mu_{D,N}=\mathbb E[D\mid o]$ and $\mathbb E[D]=p_{\rm o}\mu_{D,N}$, since $D=0$ on each even source. At 24 sites the odd and all-background mean absolute depth corrections ($10^{-3}$) are approximately $+0.586$ and $+0.276$ at the narrow width, and $+1.066$ and $+0.503$ at the broad width. The odd means use the exact conditional orbit probabilities. On 6,144 sites $\bar D_o$ denotes the mean of the 170 realized odd sources. The more precise printed bounds for normalization before averaging in Table~S6 are $\bar D_o/4\in[-0.9767,-0.8567]\times10^{-3}$ at the narrow width and $[-1.3953,-1.2656]\times10^{-3}$ at the broad width in absolute spin units. These bounds retain the frequencies' shared numerical uncertainty. The six annotations in Fig.~3 retain tighter numerical bounds than can be recovered from the separately rounded Table~S1 ordinates. Table~S6 gives the four odd-source corrections. For the complete $N=24$ background distribution, conservative propagation of the displayed Table~S1 entries gives $[0.2745,0.2785]$ and $[0.5000,0.5040]$ in $10^{-3}$ absolute spin units, compatible with the retained annotations $[0.2763,0.2764]$ and $[0.5027,0.5029]$. The accompanying reduction recovers both complete-$N=24$ annotations from the exact orbit probabilities and original eligible-source depth boxes, with exactly zero corrections on even sources. The paired medians in Table~S2 require a separate order-statistic calculation.

\textbf{Complete 6,144-site cohort.} Since the other 150 assignments have $D=0$ exactly,
\[
 \bar D_{\rm all}=\frac{170}{320}\bar D_o=\frac{17}{32}\bar D_o.
\]
The following reduction multiplies the printed Table~S6 endpoints by $17/32$ exactly and rounds the resulting bounds outwards to four decimal places. All entries are in $10^{-3}$ absolute spin units.
\begin{center}
\begin{tabular}{@{}ccc@{}}\toprule
$\gamma/\Js$ & Printed odd-source mean & Complete 320-source mean\\\midrule
\CohortRows
\bottomrule\end{tabular}
\end{center}
These are numerical bounds on the realized cohort, not confidence intervals for the all-source disorder-population mean. Both are negative, whereas the exact 24-site all-source means quoted above are positive. No even-source spectra are needed for this identity. It supplies the complete-cohort correction but does not reconstruct the missing full large-system absolute spectra. The all-source medians remain zero as shown in Sec.~S2; the primary paired-median finding remains conditional on odd eligibility.

\par\medskip
\noindent\begin{minipage}{\linewidth}
\pdfbookmark[2]{Table S2. Paired conditional medians}{table-s2}
\textbf{Table S2. Paired conditional medians.} Entries are in $10^{-3}$ of the unit-normalized total intensity. The two exact-probability rows have numerical enclosures only. Other entries are the protected fresh population-median boxes. $m$ is the eligible count out of 320.
\begin{center}
\resizebox{\linewidth}{!}{\begin{tabular}{@{}lcccccc@{}}\toprule
Sites & $\gamma$ & $m$ & $X(1.8)$ & $X(2)$ & $X(2.2)$ & $D$\\\midrule
\MedianRows
\bottomrule\end{tabular}}
\end{center}
\end{minipage}

\textbf{Central signs across the size sequence.} On the 24-site torus the exact conditional median $D$ is positive, about $0.001408$ at $\gamma=0.05\Js$ and $0.003171$ at $\gamma=0.10\Js$, in units of total spin intensity. The central intensity decreases in the paired median at all three central energies and both filter widths. At $c=2\Js$ the median $X$ is approximately $-0.0000950$ and $-0.0003911$, respectively. A negative central change can accompany shallowing because the shoulders can lose more intensity than the center.

At the largest size all three median $X$ values are positive at both filter widths (Table~S2). Across the twelve sampled geometry/filter-width settings, 12 resolve positive $X(1.8)$, 11 positive $X(2)$, 10 positive $X(2.2)$, and nine negative $D$. The three unresolved central signs all occur on 96 sites. These statements concern specified discrete energies. They require neither uniform enhancement across the full band nor a two-sided trough on every background.

\noindent\begin{minipage}{\linewidth}
\textbf{Table S3. Prespecified differences of population medians.} Entries are in $10^{-3}$ normalized units. The first two rows are successive sizes and the last row is the same-area shape comparison. Subtracting simultaneous boxes $[L_B,U_B]-[L_A,U_A]$ gives $[L_B-U_A,U_B-L_A]$. Geometry samples have independent IDs. Every listed difference admits zero.
\begin{center}
\resizebox{\linewidth}{!}{\begin{tabular}{@{}lcccc@{}}\toprule
Comparison & $X(2)$, .05 & $D$, .05 & $X(2)$, .10 & $D$, .10\\\midrule
\StabilityRows
\bottomrule\end{tabular}}
\end{center}
\end{minipage}

The tolerance ladder in normalized units is $0.0005,0.001,0.002$. With the narrower filter, only the loosest member contains the four $X(2)$ difference boxes. No listed depth difference or broad-filter central difference fits in the largest tolerance band. The fits in Table~S4 address the odd-source mean, a different target from the median differences in Table~S3. A sign or equivalence statement at fixed filter half-width does not specify a limit as $\gamma\to0$.

\par\medskip
\noindent\begin{minipage}{\linewidth}
\subsection*{Disorder means and sensitivity of the extrapolated intercept}

\textbf{Table S4. Odd-source mean correction to the mean-shoulder depth at finite size and its fitted intercept.}
All entries are in $10^{-3}$ of the unit-normalized total intensity. Divide by four for absolute spin units. For each filter setting the first column gives a numerical enclosure for the realized cohort mean (upper table) or its linear fitted intercept (lower table). The second column gives the published approximate, pointwise 95\% bootstrap interval for the corresponding population-mean quantity, including numerical uncertainty.
\begin{center}\small
\begin{tabular}{@{}l cc cc@{}}\toprule
 & \multicolumn{2}{c}{$\gamma/\Js=0.05$} & \multicolumn{2}{c}{$\gamma/\Js=0.10$}\\
\cmidrule(lr){2-3}\cmidrule(l){4-5}
Sites & Fixed mean & Pointwise & Fixed mean & Pointwise\\\midrule
\MeanDepthRows
\bottomrule\end{tabular}

\medskip
\resizebox{\linewidth}{!}{\begin{tabular}{@{}ll cc cc@{}}\toprule
 & & \multicolumn{2}{c}{$\gamma/\Js=0.05$} & \multicolumn{2}{c}{$\gamma/\Js=0.10$}\\
\cmidrule(lr){3-4}\cmidrule(l){5-6}
Fit sizes & Predictor & Fixed intercept & Pointwise & Fixed intercept & Pointwise\\\midrule
\FittedDepthRows
\bottomrule\end{tabular}}
\end{center}
\end{minipage}

For each nonzero $\gamma$, the archived unweighted fits use $\mu_{D,N}=a+b/L$ and $\mu_{D,N}=a+b/L^2$, with size index $L=\sqrt N$. The four-size window includes homothetic $N=384,1536,3456,6144$. The three-size window omits 384. Both exclude the rectangle. The intercept is a linear combination of the finite-torus means. The bootstrap resamples whole eligible backgrounds within each geometry and uses independent resamples across geometries. For each draw the numerical lower and upper endpoints propagate with the signs of the fit coefficients. The 2.5th and 97.5th percentiles give the reported pointwise envelope. The four fits share observations.

All four narrow-width intercept intervals lie strictly below zero. With the broader filter the three-largest-size $1/L$ bootstrap upper endpoint is $+0.179\times10^{-3}$, although its fitted fixed-cohort interval is wholly negative. This example distinguishes uncertainty across disorder draws from numerical error on each draw. Neither ansatz has a derived exponent or a controlled remainder for this observable. The intervals quantify sensitivity within the specified fits. They provide no coverage for uncertainty in the assumed finite-size scaling form, the zero-filter-width limit, or the interacting magnet.

\textbf{Magnitude stability and mean extrapolations.} 

We tested finite-size stability of the conditional population medians by requiring the \emph{entire} signed difference interval to lie within a specified tolerance. For both successive homothetic changes, $1536\to3456$ and $3456\to6144$, equivalence remains unresolved at the main absolute resolutions of $0.001$ for $X(2)$ and $0.002$ for $D$. The equal-area shape comparison likewise remains unresolved at both principal resolutions and both filter widths.

There is a weaker quantitative statement. At $\gamma=0.05\Js$, the two successive $X(2)$ differences have absolute upper bounds $0.001621$ and $0.001821$. The full $1536\to6144$ change and the same-area shape difference have bounds $0.001936$ and $0.001878$. All four therefore fit inside $\pm0.002$. This allowance is comparable with the narrow central effect itself. Broad-filter $X(2)$ and the depth differences with either filter fail to establish even this equivalence. The signed difference intervals include zero, leaving room both for further drift and for smaller changes. Table~S3 gives the complete median comparison.

For the \emph{odd-source population mean} $D$ on 6,144 sites, the approximate pointwise bootstrap intervals in normalized units are $[-5.504,-1.824]\times10^{-3}$ at $\gamma/\Js=0.05$ and $[-8.217,-2.390]\times10^{-3}$ at $0.10$. All four tested extrapolations have pointwise bootstrap intervals wholly below zero for the narrow-filter intercept, while one of the four broader-filter intervals includes zero (Table~S4). These fits reuse finite tori and have no derived size-scaling relation for this observable.

\par\medskip
\noindent\begin{minipage}{\linewidth}
\subsection*{Which sector changes the shoulder contrast?}

\textbf{Table S5. Signed accounting on the odd sources at $\gamma=0.05\Js$.}
For each background $i$ let $m_i=1-w_{1,i}$, $h_i$ be the filtered intensity supplied by all $q\ge3$ sectors, and $\ell_i=m_iP_i$ be the one-particle intensity removed when the rescaling is undone. Then $X_i=h_i-\ell_i$ and $D_i=C(X_i)$, where $C(f)=f(2)-[f(1)+f(3)]/2$. Positive $D$ denotes shallowing of the defined mean-shoulder contrast. The mean higher-sector mass rises from $0.12105$ on 24 sites to $0.49080$ in the large odd-source cohort. Its filtered contribution also becomes more shoulder weighted: $C(h)$ changes from positive to negative in the mean. The net correction depends on the balance between this added intensity and the removed normalization weight $\ell$, not on the higher-sector mass alone. The exact mean weights the 3,892 eligible 24-site orbits by their conditional probabilities. The fixed-cohort mean weights the 170 eligible 6,144-site draws equally. Entries are numerical interval midpoints in $10^{-3}$ absolute spin units, rounded to $0.001$. The largest numerical-enclosure half-width is below $0.060$ in these units.
\begin{center}\small
\begin{tabular}{@{}lrrrrrr@{}}\toprule
&\multicolumn{3}{c}{24 sites, exact odd mean}&\multicolumn{3}{c}{6,144 sites, fixed odd cohort}\\
\cmidrule(lr){2-4}\cmidrule(l){5-7}
 & Added $h$ & Removed $\ell$ & Net $X$ & Added $h$ & Removed $\ell$ & Net $X$\\\midrule
Left shoulder, $\Js$ & 0.049&1.059&$-1.010$&3.037&2.010&$+1.027$\\
Center, $2\Js$ &0.283&0.270&$+0.013$&1.560&1.080&$+0.480$\\
Right shoulder, $3\Js$ &0.417&0.554&$-0.136$&3.076&1.311&$+1.766$\\\midrule
Center minus mean shoulder & $+0.050$&$-0.536$&$+0.586$&$-1.497$&$-0.580$&$-0.917$\\
\bottomrule\end{tabular}
\end{center}
\end{minipage}

\textbf{What changes on individual backgrounds.} Ranking the paired $D_i$ gives a 24-site conditional median of approximately $+0.352$ and a 6,144-site fixed-cohort median in $[-0.789,-0.675]$ in Table~S5 units. The median $C(h_i)$ is slightly negative even on the small torus, where its mean is positive. The median of $-C(\ell_i)$ is positive at both sizes. The reversal comes from how the two terms compare on the same orbit. Both favor the shoulders and the added excess exceeds the removed excess, $-C(h_i)>-C(\ell_i)>0$, on 7.615\% of the exact 24-site odd probability and on 89--94 of the 170 large draws. The net change at \emph{each} shoulder exceeds the center change on 20.28\% and 98--101/170, respectively. Each count range runs from definite to possible after propagation of the saved numerical intervals.

\textbf{Sector normalization and source composition.} On each eligible source $D_i/m_i=C(h_i/m_i)-C(P_i)$ compares two unit-mass sector shapes. Averaging $D_i/m_i$ gives $+0.01889$ on 24 sites and $[-0.006960,-0.005806]$ on 6,144 sites in unit-normalized units. At both filter widths this mean is negative on every fresh geometry from 384 sites onward, including both 1,536-site shapes. At 96 sites it is positive although the mean $D_i$ is negative. The 24-site mean center loses a small amount with the broader filter, even as the mean-shoulder depth still decreases.

To hold the local flux composition fixed, group the odd sources by the number of negative pre-insertion plaquettes among the two flipped by the spin. The exact 24-site shares for zero, one, and two are 27.86, 49.44, and 22.70\%. The 6,144-site counts are 47, 78, and 45. Giving the large-cohort classes the small-torus shares leaves the large mean $D$ in $[-3.950,-3.469]\times10^{-3}$ in unit-normalized units. The paired median also stays negative. The mixed-sign class accounts for most of the shift at common shares, and its own paired median changes sign.

A separate reweighting gives the four equal-width bins of the one-particle mass $w_1$ their exact 24-site shares. The band $w_1\ge0.75$ carries 90.64\% of the small-torus odd probability and contains only 26 of the 170 large draws. The reweighted large mean $D$ spans $[-0.005,+0.364]\times10^{-3}$ with the narrower filter and $[+0.551,+0.955]\times10^{-3}$ with the broader filter in normalized units. Thus matching these four one-particle-mass-bin proportions removes the resolved narrow-filter \emph{mean} reversal and changes the broad-filter mean sign. This is a sensitivity of the ensemble mean to source composition; it is distinct from the primary paired-median result under the original background distribution conditioned on odd eligibility. The reweighted narrow paired median permits either sign. The mass $w_1$ is itself a quench outcome, and each band retains variation in the member spectra. Giving entire backgrounds new weights changes the averaging population; it is not automatically a correction to the original ensemble. The association with $w_1$ neither identifies a microscopic cause nor establishes a universal reversal at fixed composition.

The exact 24-site orbit records combine $q\ge3$ and lack separate values of $w_3$ and the filtered $I_3(c)$. Hence the matched one-plus-three-particle response $(w_1P+I_3)/(w_1+w_3)$ cannot be compared on the 24-site and 6,144-site tori. The accounting identifies the spectral contribution responsible for the observed mean-shoulder correction reversal. Its microscopic origin and its infinite-size sign remain open.

\par\medskip
\noindent\begin{minipage}{\linewidth}
\pdfbookmark[2]{Table S6. Normalization of the 24-site and 6,144-site means}{table-s6}
\textbf{Table S6. Normalization before and after the odd-source average.}
The entries are numerical enclosures of $d(P)-d(\langle S_b\rangle_o)$ in $10^{-3}$ absolute local spin intensity ($S^z=\sigma^z/2$). The same full spectrum enters both columns. Positive values mean that restoring higher sectors reduces the mean-shoulder depth. At 24 sites the average uses exact conditional orbit probabilities. At 6,144 sites it uses all 170 realized eligible assignments with equal weights.

\begin{center}
\begin{tabular}{@{}rrrcc@{}}\toprule
$N$ & $\gamma/\Js$ & Odd sources & Normalize each source & Normalize after averaging \\\midrule
\NormalizationRows
\bottomrule\end{tabular}
\end{center}
\end{minipage}

For source $b$, let $w_b$ be its one-particle mass and $P_b$ its one-particle spectrum normalized to unit mass. Before rescaling, that sector contributes $w_b P_b$. Writing $\langle\cdot\rangle_o$ for the same conditional average in every term gives
\[
 P_{\rm each}=\langle P_b\rangle_o,\qquad
 P_{\rm pool}=\frac{\langle w_b P_b\rangle_o}{\langle w_b\rangle_o},\qquad
 d(U)=\frac{U(1)+U(3)}2-U(2).
\]
Linearity gives $d(P_{\rm pool})=\langle w_b\,d(P_b)\rangle_o/\langle w_b\rangle_o$. The calculation encloses each product with its own mass and three prespecified ordinates. On the 6,144-site torus it intersects these bounds with the directly saved signed kernels for the full, matched, and unnormalized one-particle responses. The numerator is divided by a single positive mean-mass enclosure. The column for normalization before averaging also uses the directly saved paired $D_b$ bounds. Interval arithmetic preserves the shared-input enclosures without an independence assumption. The conversion from unit-mass to absolute spin intensity is $1/4$.

Both prescriptions give shallowing on 24 sites and deepening on 6,144 sites at both filter widths. Pooling gives greater weight to sources with larger $w_b$. This retrospective sensitivity check describes the exact small-torus probability distribution and the fixed large cohort. Its numerical intervals carry no disorder-population coverage for the large mean. A common normalization of the averaged spectrum defines no new paired median.

\subsection*{Relative error in the conditional mean contrast}

When the one-particle control is normalized separately for each source, the relative error in the conditional mean contrast is
\[
 R_{N,\gamma}=\frac{M_{N,\gamma}}{F_{N,\gamma}},\qquad
 M_{N,\gamma}=\langle D_\gamma\rangle_o,\qquad
 F_{N,\gamma}=d(\langle S_\gamma\rangle_o).
\]
The numerator and denominator use the same weighting, conditional on odd eligibility: exact probabilities over the 3892 eligible $N=24$ orbits, or equal weights on the 170 realized eligible $N=6144$ draws. The control is normalized before averaging, as in the corresponding column of Table~S6. Dividing both $M$ and $F$ by four converts them to absolute spin units without changing $R$.

\noindent\begin{minipage}{\linewidth}
\pdfbookmark[2]{Table S7. Relative mean-contrast error}{table-s7}
\textbf{Table S7. Relative mean-contrast error with normalization before averaging.}
The displayed $M/4$ and $F/4$ boxes are in $10^{-3}$ absolute spin units and use printed Tables~S6 and S1, respectively. Both percentage columns round outwards to 0.1 percentage point. The last column instead reduces the original saved records described below. Positive values mean that the control overestimates the full mean contrast; negative values mean that it underestimates it.
\begin{center}\small
\begin{tabular}{@{}rrcccc@{}}\toprule
$N$ & $\gamma/\Js$ & Printed $10^3 M/4$ & Printed $10^3 F/4$ & $R$ (printed) & $R$ (original)\\\midrule
24 & 0.05 & $[0.5861,0.5862]$ & $[3.725,3.727]$ & $[15.7\%,15.8\%]$ & $[15.7\%,15.8\%]$\\
24 & 0.10 & $[1.0663,1.0665]$ & $[6.101,6.103]$ & $[17.4\%,17.5\%]$ & $[17.4\%,17.5\%]$\\
6144 & 0.05 & $[-0.9767,-0.8567]$ & $[1.7510,1.8805]$ & $[-55.8\%,-45.5\%]$ & $[-55.7\%,-45.6\%]$\\
6144 & 0.10 & $[-1.3953,-1.2656]$ & $[3.6435,3.7855]$ & $[-38.3\%,-33.4\%]$ & $[-38.3\%,-33.4\%]$\\
\bottomrule\end{tabular}
\end{center}
\end{minipage}

For full-ordinate boxes $[L_j,U_j]$ at $j=1,2,3$, the conservative full-contrast interval is
\[
 F\in\left[\frac{L_1+L_3}{2}-U_2,\ \frac{U_1+U_3}{2}-L_2\right].
\]
All four lower bounds are strictly positive. The quotient enclosure is the minimum and maximum of the four endpoint quotients $M_i/F_j$. This interval operation remains conservative when the numerator and denominator share uncertainty; it assumes no independence and adds no new confidence guarantee. For example, the large-torus printed box for the narrower filter gives $R\in[(-0.9767)/1.7510,\,(-0.8567)/1.8805]$, or $R\in[-55.8\%,-45.5\%]$ after outward rounding.

The original-record calculation averages the saved paired $D$ boxes for $M$. At $N=24$ it also combines the three full-ordinate boxes under the exact conditional orbit weights. At $N=6144$ the full-response signed kernel directly encloses $S(2)-[S(1)+S(3)]/2=-d(S)$; negating its averaged box gives $F$. This retains the shared-frequency numerical uncertainty. All eligible rows, including wider numerical boxes, remain included. The original input and ratio enclosures lie inside the conservative printed-input bounds. The small narrowing visible at $N=6144$, $\gamma/\Js=0.05$, reflects original precision and the direct signed kernel, not a changed population or normalization.

The ratio compares conditional mean contrasts; the averages are taken before division. Its large-torus bounds describe the realized odd cohort and enclose numerical uncertainty, rather than disorder-population uncertainty. The full conditional mean contrast remains positive, while its approximation error changes sign. This complementary mean-based diagnostic is distinct from the primary paired-median reversal and from an error measure for the full spectrum. A positive mean-shoulder contrast alone does not establish a two-sided trough.

\medskip
\textbf{From a saved source to Table S2.} At the narrower filter setting, sorting 24-site lower $D_b$ endpoints by value places odd orbit 2212 at the half-probability crossing. Its lower endpoint is $0.00140793856\ldots$ in unit-normalized intensity. The cumulative odd probability rises from $0.49974337$ to $0.50004863$ there. Sorting upper endpoints separately gives orbit 4508 at $0.001408058546\ldots$. Rounding those boundaries outwards gives $[1.407,1.409]$ in the $10^{-3}$ normalized units of Table~S2. The accompanying exact-probability reduction reproduces these median boundaries from the supplied orbit catalog, probability-class counts, and original source boxes. All 8192 orbit weights sum exactly to one, and eligibility is retained for all orbits; only the 3892 odd sources enter the conditional median.

\textbf{Table reductions.} The accompanying scripts apply the stated probability weights, signed-kernel combinations, and order-statistic rule to saved numerical enclosures. They reproduce all 48 fresh-cohort median intervals and all eight exact-probability median intervals in Table~S2, retaining the wider rows, and the four Table~S6 rows for both normalization prescriptions. Section~S2 describes validation of the spectral calculations; Sec.~S6 maps the supplied inputs and scripts to the reported reductions.

\par\medskip
\secS{4}{Two-filter support bounds and per-background decisions}

\begin{figure}[t]
\includegraphics[width=.96\linewidth]{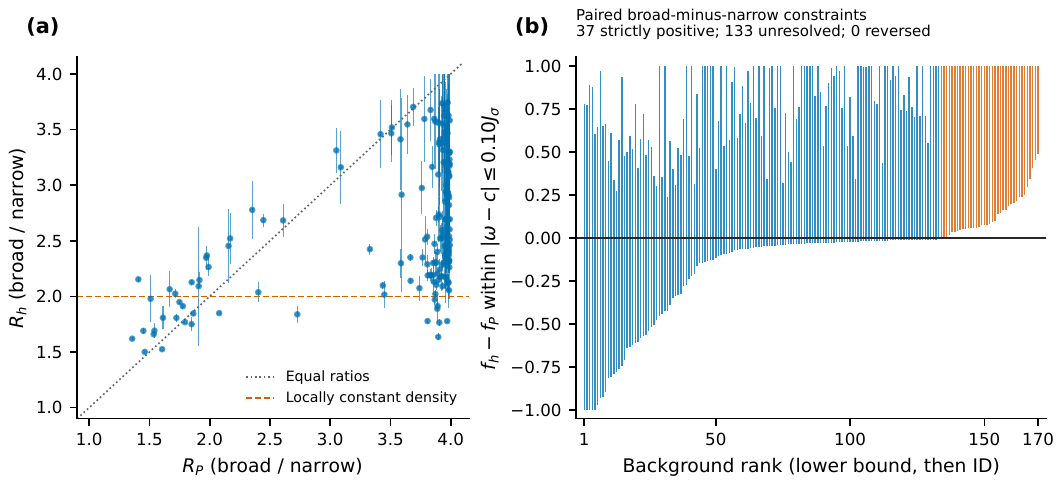}
\caption{\WidthCaption}\label{fig:width}
\end{figure}

\subsection*{Pairing the two filter widths}

In Eq.~(6) of the article, the terms entering the central correction are positive quantities. At $N=6144$ the fixed-cohort mean of $h$ at $2\Js$ changes from $[0.0061168,0.0063616]$ to $[0.0129734,0.0132432]$ when the filter half-width doubles. The corresponding mean of $(1-w_1)P$ changes from $[0.0043201,0.0043211]$ to $[0.0088111,0.0088118]$. Both roughly double. The ratio of two means weights each individual ratio by its narrow-filter contribution. Sources with a dark one-particle window then receive little weight.

A more direct diagnostic preserves each background pairing. For $A=P$ or $h$ put $R_A=A_{2\gamma}/A_\gamma$, with fixed $c$ and $\gamma=0.05\Js$. The factor $1-w_1$ cancels in the ratio of the one-particle matching term. The fixed-cohort median ratio at $2\Js$ lies in $[3.908,3.913]$ for $P$ and $[2.314,2.645]$ for $h$. Of the 170 pairs, 118 resolve $R_h<R_P$, 22 resolve the reverse, and 30 overlap [Fig.~\ref{fig:width}(a)]. At $1.8\Js$ and $2.2\Js$, 127 and 113 pairs, respectively, resolve the same lower higher-particle ratio. These are descriptive row comparisons.

For a single positive line at detuning $d=|\omega-c|$, the broad-to-narrow ratio is
\begin{equation}
 R(d)=4\,\frac{d^2+\gamma^2}{d^2+4\gamma^2}.
 \label{eq:ratio}
\end{equation}
It increases from one on resonance to four for distant lines. For a positive spectral measure the measured ratio is the average of Eq.~(\ref{eq:ratio}) under weights proportional to its \emph{narrow-filtered} intensity. Thus the near-four one-particle ratios on many backgrounds directly indicate tail weighting. A smooth, locally constant density integrated against a peak-height-one Lorentzian contributes in proportion to its area $\pi\gamma$, giving a ratio near two when remote tails are controlled. A sparse spectrum can produce the same result. A single line at $d=\sqrt{2}\gamma$ gives exactly two. The expected distinction between locally smooth higher-particle weight and distant one-particle tails can therefore be probed by the two filter widths, but neither smoothness nor a hard one-particle hole follows from a ratio alone.

\subsection*{A bound on the contribution from the central window}

A lower broad-to-narrow response ratio need not imply a greater fraction inside any particular integration window. We ask this stricter question at $c=2\Js$ using \emph{all} 170 eligible $N=6144$ backgrounds. For each channel define
\begin{equation}
 f_A=\frac{\int_{|\omega-c|\le 2\gamma}F_\gamma(\omega-c)\,d\mu_A}
           {\int F_\gamma(\omega-c)\,d\mu_A},\qquad A=P,h .
 \label{eq:f}
\end{equation}
Here $d\mu_P=d\mu_1/w_1$ and $d\mu_h$ is the full combined higher-particle measure. Each fraction refers to its own narrow-filter signal. The half-width of the \emph{hard integration window} is $2\gamma=0.10\Js$.

Set $R_*=R(2\gamma)=5/2$. On the inside, $1\le R(d)\le R_*$. Outside, $R_*\le R(d)<4$. Combining these ranges gives, conservatively,
\begin{equation}
 R_*-(R_*-1)f_A \ \le\ R_A\ \le\ 4-(4-R_*)f_A .
 \label{eq:twosides}
\end{equation}
For an observed enclosure $[R_A^-,R_A^+]$, it follows that
\begin{align}
 L_A&=\max\!\left(0,\frac{R_*-R_A^+}{R_*-1}\right),\nonumber\\
 U_A&=\min\!\left(1,\frac{4-R_A^-}{4-R_*}\right),\qquad L_A\le f_A\le U_A .
 \label{eq:bounds}
\end{align}
The inside fraction for $h$ is forced to exceed that for $P$ only when $L_h>U_P$ under this bound comparison. The reverse ordering requires $L_P>U_h$.

The first inequality holds on \textbf{37 backgrounds}, the reverse on \textbf{zero}, and the two sets of allowed fractions overlap on the other \textbf{133}. Figure~\ref{fig:width}(b) shows the corresponding enclosures for $f_h-f_P$. This descriptive result uses published positive ordinates and their paired differences between the two filter settings, with numerical uncertainty retained. It is not a separate population inference. The absence of a forced reverse case leaves the sign on each of the 133 overlapping cases open.

The one-particle boxes also give a 6.13\% upper bound on the fixed-cohort median share of the narrow-filter one-particle signal arising inside the window. Even a small filtered contribution can include a line inside the window. The scalar evidence rules out a strict one-particle hole of this radius on at least 29 backgrounds. For at least 75 backgrounds, the higher-particle signal cannot come entirely from outside the window. Proving that some higher-particle intensity is inside is weaker than proving that its inside fraction exceeds the one-particle fraction. The 37 strict cases quantify the latter distinction without reconstructing a line spectrum.

\subsection*{Correlated numerical bounds and retained decisions}

At $N=6144$ and $c=2\Js$, take $A=P$ or $h$ on one eligible ID. The published bounds on the narrow term $A_n$, the broad term $A_b$, and the difference $A_b-A_n$ define a polygon of feasible pairs. All 170 narrow denominators have positive lower bounds. The ratio $A_b/A_n$ is monotone or constant along each edge, so its extrema occur at vertices. We evaluate it at every feasible vertex, intersect its range with $[1,4]$, and round outwards only for export. The reduction uses exact rational arithmetic on the recorded decimal inputs.

The bounds in Eqs.~(\ref{eq:twosides}) and (\ref{eq:bounds}) use the narrow-filter-weighted measure $d\pi_A=F_\gamma\,d\mu_A/A_n$ and the window $|\omega-c|\le2\gamma$. Allowing the boundary ratio $5/2$ on either side is conservative. For each ID the difference box is $[L_h-U_P,U_h-L_P]$: a positive lower endpoint forces $f_h>f_P$, a negative upper endpoint forces the reverse, and contact with zero remains unresolved. Changing from $P$ to the unscaled $q=1$ measure cancels in the fraction; replacing its denominator by unfiltered sector mass or by the combined spin signal would define a different observable. The same applies to $h$, including its bounded higher-sector tail. These are sufficient bounds, not an inversion of the spectrum.

\textbf{All 170 retained rows.} The strict counts are 37 higher, 0 one-particle, and 133 unresolved. Adding the published paired-difference constraints to the recorded broad/narrow quotient boxes changes none of the classifications. Among the higher-fraction cases, the first ID in numerical order is 003. It has $R_P\in[3.9301,3.9330]$ and $R_h\in[2.0599,2.1227]$. The marginal interval for $f_P$ reaches at most 0.04657 and the interval for $f_h$ starts above 0.25157. The 37 IDs with forced higher-particle ordering are
\par\smallskip
\texttt{003}, \texttt{017}, \texttt{018}, \texttt{029}, \texttt{034}, \texttt{064}, \texttt{070}, \texttt{071}, \texttt{090}, \texttt{091}, \texttt{100}, \texttt{106}, \texttt{112}, \texttt{123}, \texttt{125}, \texttt{126}, \texttt{132}, \texttt{139}, \texttt{149}, \texttt{152}, \texttt{161}, \texttt{164}, \texttt{176}, \texttt{179}, \texttt{181}, \texttt{192}, \texttt{208}, \texttt{216}, \texttt{242}, \texttt{252}, \texttt{253}, \texttt{259}, \texttt{277}, \texttt{286}, \texttt{289}, \texttt{292}, \texttt{311}.

\par\smallskip
The accompanying data include all 170 per-ID window-fraction rows. Classifying their saved difference bounds reproduces the $37/0/133$ counts and the listed forced IDs. Figure~\ref{fig:width}(a) shows the marginal broad/narrow quotients; Fig.~\ref{fig:width}(b) shows the bounds including the paired-difference constraints. The $118/22/30$ ordering counts compare broad-to-narrow filtered-intensity \emph{ratios}. They are distinct from the $37/0/133$ window-\emph{fraction} ordering.

\textbf{Prospective central-sign test on 384 sites.} A separate test asked whether an equal-time moment could certify enhancement on one fresh odd source. The first ordered draw (000) qualified. At $c=2\Js$ and $\gamma=0.05\Js$, the mean squared detuning in the higher sectors is $M_{2,h}=\int(\omega-c)^2\,d\mu_h/(1-w_1)$. Jensen's inequality gives $h/(1-w_1)\ge B=\gamma^2/(\gamma^2+M_{2,h})$. Since $X=(1-w_1)[h/(1-w_1)-P]$, a certified $B>P$ would predict $X>0$. The matched ordinate $P\simeq0.15165220$ requires $M_{2,h}<\gamma^2(P^{-1}-1)\simeq0.0139851\Js^2$ for this prediction. The local Wick calculation instead gave $M_{2,h}\simeq16.329024\Js^2$ and $B\simeq0.00015307818$. The quantity $B-P$ is a lower bound on $X/(1-w_1)$. Its entire numerical enclosure lay below $-0.15149911$, so this test yielded no sign for $X$. A remote line enters the moment with its squared detuning, so even a precise moment can leave the central intensity undetermined. The test stopped at that point, before any full-response ordinate was evaluated.

\par\medskip
\section*{S5.\quad Physical spin response and elastic subtraction}
\addcontentsline{toc}{section}{S5. Physical spin response and elastic subtraction}

On the tilted 24-site torus the physical Hamiltonian is Eq.~(8) of the article with $J=1/2$, $b=1/4$ or $11/40$, and $S=\sigma/2$. The exchange scale multiplying Pauli products is $\Js=1/8$, and the field coefficient multiplying each Cartesian Pauli component is $-b/(2\sqrt{3})$. These two fields have $b/J=0.50,0.55$. The frozen $\lambda$ and independent-link distribution do not define a map from $b$ to an effective Hamiltonian at each field.

\textbf{Scope of the frozen benchmark.} The frozen preparation is a stationary mixture of allowed sector ground states, not a global ground state of the interacting magnet or of one common frozen Hamiltonian. Its negative-transfer spin mass is genuine, independent of the Lorentzian tails of positive-transfer lines. The comparison below tests the positive-frequency predictions of this prespecified frozen ensemble against the interacting reference.

\textbf{The prediction at $b/J=0.50$.} Both all-orbit frozen spectra were fixed before the physical band measurements entered the comparison. They include signed and zero transfers and use the same uncentered local spin probe as the physical calculation. The raw physical intensities were obtained for the baseline numerical state. The same state later received numerical rank-one ground-space qualification. For $(c/\Js)=1,2,3$ and $\gamma/\Js=0.05$, the absolute physical boxes, multiplied by $10^3$, are $[6.6399,6.6516]$, $[3.7541,3.7673]$, and $[3.6010,3.6159]$. Figure~4 of the article uses the model prediction fixed before the benchmark and the original raw physical intensities. The corresponding frozen all-ensemble values at this filter setting can also be read from Table~S1.

The matched and full squared errors are $[4.72632,4.82323]\times10^{-6}$ and $[3.33607,3.41990]\times10^{-6}$ in squared absolute-intensity units. The signed gain can be evaluated with the common physical ordinate as
\[
 G=\sum_j(\absI_{S,j}-\absI_{P,j})(2Y_j-\absI_{S,j}-\absI_{P,j}),
 \qquad \absI_{S,j}=S_j/4,\quad \absI_{P,j}=P_j/4 .
\]
The center-specific gains in order $1,2,3$ are $[1.22269,1.23372]\times10^{-6}$, $[1.13449,1.15220]\times10^{-8}$, \newline and \mbox{$[1.56185,1.58111]\times10^{-7}$}. The endpoints round the archived paired bounds outwards. The approximately 29\% reduction in the three-frequency squared error is dominated by the left shoulder; the full prediction retains the nonzero squared residual stated above and the incorrect right-shoulder ordering.

\textbf{Benchmark residual and insertion cost.} With equal sector energy offsets, approximately 13.72\% of its integrated spin intensity has strictly negative transfer before broadening. The populated sectors permit downward transitions that a zero-temperature ground state forbids. Figure~4 of the article shows the raw absolute filtered physical intensity $Y$ and the two predictions. With $c_j/\Js=1,2,3$, define the unweighted three-band error gain
\begin{equation}
 G=\sum_j (Y_j-\absI_{P,j})^2-\sum_j(Y_j-\absI_{S,j})^2.
 \label{eq:gain}
\end{equation}
Restoring the odd-source multiparticle response reduces the squared error by 28.8--29.7\%. Of the gain $G\in[1.39022,1.40336]\times10^{-6}$ in squared absolute-intensity units, $[1.22269,1.23372]\times10^{-6}$ comes from the left shoulder. Both frozen spectra nevertheless place the right shoulder above the center. The full model gives a right-minus-center difference of $+0.0019624$ in absolute spin intensity, while the physical interval is $[-0.00016619,-0.00013825]$. The center itself changes by only $6.17\times10^{-6}$ and reaches 75.3--75.6\% of its measured intensity.

The residual norm of the full prediction is 3.80--3.85 times the length of the correction to the three predicted ordinates. Thus the left-shoulder improvement leaves the wrong right-shoulder ordering and a substantial residual at the specified field.

The ground-state limitation is also reflected in the insertion cost, which is too high for either adjacent-face sign by about ten narrow-filter half-widths. We next test whether changing the frozen flux populations improves the comparison.

\textbf{Measured pair populations and a restricted reweighting.} Before insertion, write $s=W_0$ and $t=W_{11}$ for the flux signs of the two plaquettes sharing the flipped bond. The four measured pair populations are the probabilities of $(s,t)=(++,+-,-+,--)$ in the interacting baseline state. In the baseline numerical state at $b/J=0.50$ the face means agree within their numerical intervals at $w\simeq0.139709$. For homogeneous independent links of mean $m$, $\langle W_0\rangle=m^6$ and $\langle W_0W_{11}\rangle=m^{10}=w^{5/3}$ because the shared edge cancels. Since $\Pr(s=t)=(1+\langle W_0W_{11}\rangle)/2$, the measured pair moment $0.04427011$ and the mean-matched prediction $0.03761620$ give an excess equal-sign probability of approximately $0.00332695$, or $0.333$ percentage points. The positive discrepancy survives the conditional rank-one ground-projector allowance. The intensity comparison below uses records evaluated using that same state.

\textbf{Face-0 source dephasing.} Face 0 is the adjacent plaquette with flux operator $W_0$. After spin insertion, project the source onto its $W_0=+1$ and $W_0=-1$ components and sum their responses without renormalizing either component. For a final eigenstate $|n\rangle$ and the spin-created source $|\chi\rangle$, the amplitudes from the two signs are $A_{\pm n}=\langle n|(1\pm W_0)|\chi\rangle/2$. Their coherent and dephased intensities differ by $2\,\mathrm{Re}(A_{+n}^*A_{-n})$. The filter weights multiply this difference at each signed transfer, including zero transfer for the raw response. For $O_s=\sigma_0^z$ and $O_t=W_0\sigma_0^z$, let $Q_s,Q_t$ be the filtered Pauli-source intensities. The dephased Pauli response is their average. In units of $S=\sigma/2$, the coherent intensity is $Y=Q_s/4$, the dephased intensity is $Z=(Q_s+Q_t)/8$, and their difference is $T=(Q_s-Q_t)/8$. Raw intensities retain the elastic contribution.

Let $I_g$ be the archived mean \emph{full} frozen central intensity in the ordered pre-insertion class $g=(s,t)$, in absolute spin units at $c=2\Js$, $\gamma=0.05\Js$, including the elastic contribution. A frozen initial orbit has definite face signs, so dephasing the source in the face-0 sign leaves its intensity unchanged. The archived ensemble has face mean $w_0\simeq0.05$ and class masses $p_g=[1+(s+t)w_0+st\,w_0^{5/3}]/4$. In the order $(++,+-,-+,--)$ these are approximately $(0.276697,0.248303,0.248303,0.226697)$. The measured class masses $q_g$ are approximately $(0.330922,0.238932,0.238932,0.191213)$. Replacing $p_g$ by $q_g$ while keeping the archived relative weights within each class fixes every $I_g$. Since $I_g\ge0$,
\[
 I_{\rm four}=\sum_g q_g I_g
 \ \le\ \max_g(q_g/p_g)\sum_g p_g I_g
 \ <\ 1.195976\, (0.002837554)\ <\ 0.003394 .
\]
The maximum occurs in class $++$. The raw physical face-0-dephased central intensity is at least $0.00352806096$. The unrounded bounds give a shortfall of at least $0.00013441763$ in absolute spin units. If orbit weights can vary within a class, its $I_g$ can change and this bound no longer applies. These four inputs are reported pair-and-class intervals, the fixed independent-link distribution, the previously fixed raw intensity, and the raw dephased band.

\textbf{Free relative orbit weights.} The seven-orbit fit mixes seven frozen-background gauge orbits, with their preparations, transition energies, and spin matrix elements fixed. Its nonnegative rational probabilities sum exactly to one and were fitted to the known adjacent-plaquette populations and three raw face-0-dephased spin intensities at $b/J=0.50$ and $\gamma/\Js=0.05$. This common mixture fits both face means, their product, and all four ordered pair populations. Its raw face-0-dephased intensity enclosures at $1,2,3\,\Js$ lie wholly within the intervals evaluated using that state. For this local spin probe, each occupied orbit has certified zero unbroadened negative-transfer weight $d_i$. Hence $\sum_i p_i d_i=0$ attains the minimum, since all spectral weights are nonnegative. This retrospective fit is distinct from the prespecified benchmark in Fig.~4 of the article. It shows that suitably reweighted frozen populations can reproduce these selected observables, but it does not independently predict those populations or the response. The fit does not explain the measured coherent-minus-dephased contrast. The accompanying numerical collection supplies the seven rational probabilities, the per-orbit inputs and target intervals, and a script that verifies the common mixture.

\par\medskip
\subsection*{Field displacement and flux-sign interference}
\begin{figure}[t]
\includegraphics[width=.96\linewidth]{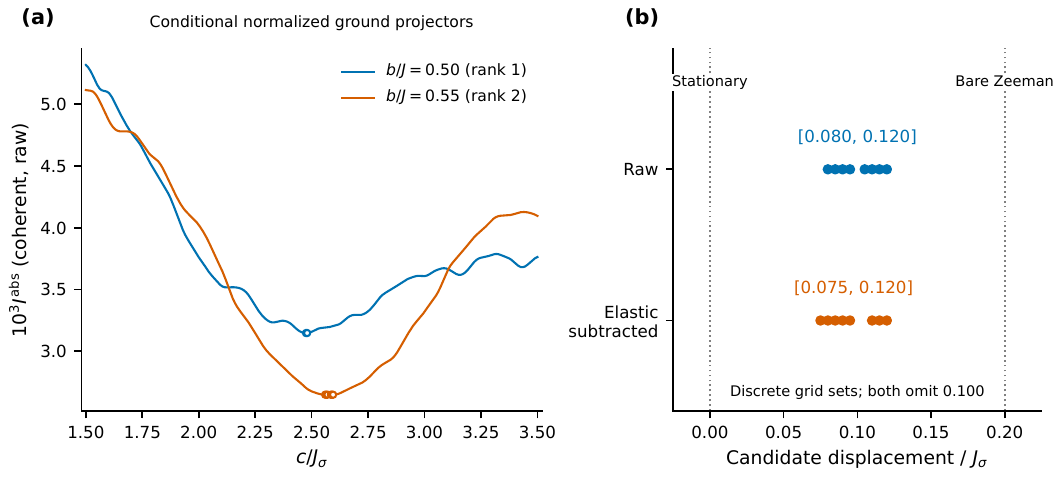}
\caption{\FieldCaption}\label{fig:field}
\end{figure}

We keep the same $b/J=0.50$ baseline and compare it with $b/J=0.55$ in Eq.~(8) of the article, using the normalized numerical ground-space projector at each field. Independent low-root evidence admits the original baseline numerical state as a rank-one projector and the higher-field preparation as a symmetry-related rank-two projector. Completeness of the identified lowest space and its separating gap remain numerical premises. The positive-filter spectra were evaluated by Chebyshev propagation. The maximum required numerical-enclosure half-width in the raw coherent intensity or its flux-interference comparison is $1.981\times10^{-7}$ in absolute spin units.

At $\gamma=0.05\Js$ the coherent signal rises by about $2.62\times10^{-4}$ at the fixed frequency $2\Js$ and falls by about $2.08\times10^{-4}$ at $2.2\Js$. The primary grid gives candidate shifts of $0.100$ and $0.125\Js$ for the raw trough. Figure~\ref{fig:field} shows the finer grid. The disconnected raw displacement set has hull $[0.080,0.120]\Js$. Subtracting the complete true-elastic weight, including off-diagonal ground-space matrix elements, gives a disconnected set with hull $[0.075,0.120]\Js$. Both sets exclude zero and the bare-Zeeman displacement $\Delta b=0.200\Js$. The value $0.100\Js$ lies in both hulls and neither fine-grid set, though the coarse grid admits it. The sets collect differences between grid minima that survive the numerical bounds at the two fields. Locating the continuous minimum would require information between grid points.

The coherent-minus-dephased comparison defined above quantifies interference between the two $W_0$ signs in the central spin response. In a retrospective comparison at fixed frequency $c=2\Js$ and filter half-width $\gamma=0.05\Js$, bare Zeeman precession gives the observed positive sign of the elastic-subtracted coherent-minus-dephased spin contrast, but overestimates its magnitude by more than a factor of 360 at $b/J=0.50$. Raising $b/J$ to 0.55 at the same probe frequency reduces the bare prediction, whereas the measured 24-site contrast increases.

The positive, sub-bare-Zeeman grid shift and the opposite intensity changes at the two fixed frequencies are empirical requirements for a field-dependent theory. Developing such a theory and determining the physical size dependence remain open questions.

\textbf{Complete elasticity and field comparison.} At each field use $\rho_b=\Pi_b/r_b$ for the normalized numerically admitted lowest-space projector. The rank-one baseline state and symmetry-related rank-two higher-field space have independent-start evidence and separated returned low roots. The conditional result relies on their representing the complete lowest manifolds. Neither the ranks alone nor the grid shifts locate a level crossing. For a normalized Pauli source operator $O$, the full elastic fraction in this preparation is
\begin{equation}
 a_O=\frac{1}{r_b}\sum_{i,j=1}^{r_b}|\langle e_j|O|e_i\rangle|^2
     =\frac{\|\Pi_b O\Pi_b\|_F^2}{r_b}.
\end{equation}
The same elastic mass, including off-diagonal matrix elements, enters at all centers with its filter value at zero transfer. Finite nonzero quasielastic weight is retained.

\textbf{Spin coherence at fixed frequency.} For $T^{\rm inel}$ we remove the full ground-space contribution from both responses defined above.

In the bare reference $H_Z=-b\sum_i S_i^{[111]}$, the ground state is polarized along $[111]$. Each Cartesian Pauli acting on one polarized spin creates a field-basis flip with probability $1-1/3=2/3$. The source $O_s$ acts on one site. In $O_t$, the face operator's $\sigma_0^y$ combines with $\sigma_0^z$ to give $i\sigma_0^x$, leaving one Pauli on each of six distinct sites. Its weight at energy $jb$ is therefore $p_j=\binom{6}{j}(2/3)^j(1/3)^{6-j}$, with $p_1=4/243$. Omitting $j=0$ and using the peak-one filter $F_\gamma(x)=\gamma^2/(x^2+\gamma^2)$ gives
\[
 T_Z^{\rm inel}=\frac{79}{972}F_\gamma(b-c)-\frac18\sum_{j=2}^6p_jF_\gamma(jb-c),
 \qquad \frac{79}{972}=\frac{2/3-p_1}{8}.
\]
At $c=2\Js$ and $\gamma=0.05\Js$, the bare one-flip line lies on resonance at $b/J=0.50$. At $0.55$ it lies four filter half-widths above the same center, so its filter factor falls from $1$ to $1/17$. The bare-reference bounds and the physical 24-site intervals give the following retrospective comparison in absolute spin units.
\begin{center}
\begin{tabular}{@{}ccc@{}}
\toprule
$b/J$ & Bare $T_Z^{\rm inel}$ & Interacting $T^{\rm inel}$ \\
\midrule
$0.50$ & $(0.08119,\,0.08128)$ & $[0.0002225385,\,0.0002228373]$ \\
$0.55$ & $(0.00472,\,0.00479)$ & $[0.0004345217,\,0.0004348554]$ \\
\bottomrule
\end{tabular}
\end{center}
The comparison used a prespecified effect size of $0.0011447/4=0.000286175$ in absolute spin units. It takes the lower endpoint of the 1,152-site frozen-source comparison (Sec.~S1) as a scale of interest. The two calculations use different transfer-energy origins. At $c=2\Js$ only the higher-field physical contrast exceeds this scale. At $2.2\Js$ both lie below it. These contrasts compare intensities at prescribed frequencies. The separate trough measurement in Fig.~\ref{fig:field} tracks the position of the minimum of $Y$ on a grid.

The primary search has spacing $0.025\Js$ over $[1.5,3.5]\Js$. It admits shifts of $\{0.100,0.125\}\Js$ for the raw response and $\{0.100\}\Js$ after elastic subtraction. The finer grid uses $0.005\Js$. A correlated signed-kernel test excludes a candidate $j$ only if a comparison $k$ certifies $Y_j-Y_k>0$, retaining shared moments, energy origin, and elastic masses. Reported displacements are all ordered higher-field minus baseline candidate differences. On the fine grid the raw baseline candidates are $2.475,2.480$ in $\Js$ and the higher-field candidates are $2.560,2.565,2.570,2.585,2.590,2.595$. In units of $\Js$, the raw fine-grid displacement candidates are $\{0.080,0.085,0.090,0.095,0.105,0.110,0.115,0.120\}$. Complete elastic subtraction gives $\{0.075,0.080,0.085,0.090,0.095,0.110,0.115,0.120\}$. Both exclude zero and $0.200$. Interpolating between endpoints of either hull would incorrectly admit $0.100$. This comparison is distinct from the three-frequency model score.

\secS{6}{Accompanying data and reproduction materials}

\textbf{AI-assisted analysis and verification.} OpenAI ChatGPT Astra and Codex configured with Astra assisted with theoretical and numerical analysis, coding, figures, and manuscript preparation under the author's direction. Recorded verification includes the finite-Fock, Gaussian, and independent-transform comparisons described in Sec.~S2 and separately implemented reductions of saved numerical bounds and exact probability weights.

\textbf{Access and capabilities.} The accompanying numerical inputs, reduction scripts, expected outputs, and selected method and validation records described below are publicly available in the \href{https://github.com/davidronquillo/kitaev-spin-spectral-contrast}{\texttt{kitaev-spin-spectral-contrast}} companion repository and are also supplied as a standalone numerical collection. Its commands run without live repository access. Editable manuscript sources and figure assets are supplied in the separate typesetting package.

The accompanying scripts reproduce the reported table reductions from the supplied numerical bounds and probability weights. Validation of the spectral calculations is described in Sec.~S2. The source manifest identifies the scientific records used for each reduction.

\begin{center}\small
\begin{tabularx}{\linewidth}{@{}>{\raggedright\arraybackslash}p{.27\linewidth}X@{}}\toprule
Claim or material & Accompanying evidence and reductions\\\midrule
Conditional paired medians (Table~S2) & All 1920 fresh source IDs, eligibility, and relevant lower/upper kernel boxes; the 8192-orbit catalog, exact probability-class counts, and 3892 eligible small-torus source boxes. Independent reduction reproduces all 48 fresh and eight exact-probability intervals, including wider numerical rows.\\[4pt]
Normalization and Figure~3 annotations & One-particle masses, depth/signed-kernel boxes, and exact small-torus weights reproduce all four Table~S6 rows for both normalization prescriptions and the two complete-$N=24$ annotations. Printed Table~S1 entries give compatible, wider bounds.\\[4pt]
Relative mean-contrast error (Table~S7) & An input snapshot and exact-rational reducer reproduce the four printed-input ratios and four original-record ratios. The original means use all 3892 exact-weight eligible $N=24$ orbits and all 170 eligible $N=6144$ draws.\\[4pt]
Seven-orbit fit (Sec.~S5) & Seven exact rational probabilities, per-orbit scalar inputs, target intervals, and a checker verify normalization, whole-interval containment for the common mixture, and zero unbroadened downward weight. The collection README gives the command and documentation.\\[4pt]
Method and validation & Selected implementation records specify the Gaussian chart, proposal, estimator, and numerical allowances. A summary of the recorded finite-Fock, high-sector, and independent-transform checks gives the comparisons in Sec.~S2.\\[4pt]
Other finite-size and physical diagnostics & Absolute-ordinate, depth/fit, window-fraction, and physical field/grid tables. Stored-value checks reproduce the window classifications, reproduce the discrete displacement sets, and verify compatibility of all 96 Table~S1 ordinate boxes.\\[4pt]
Manuscript and figures & Editable text, bibliography, numerical table sources, five quantitative vector figures, and an explanatory vector schematic, together with the schematic drawing code and the available historical plotting inputs and script. The accompanying README specifies their reproduction capabilities.\\\bottomrule
\end{tabularx}
\end{center}

The original-row reductions use the numerical endpoints, exact orbit weights, and specified rank/error-allocation rules. Their interpretation is conditional on the validity of those numerical enclosures and the sampling assumptions in Sec.~S2. The collection includes column definitions, independent reducers, and expected-output comparisons. For the relative contrast, it also supplies the input means and an extractor that reconstructs them from the source rows. A separate checker performs the printed-input arithmetic, including the $17/32$ complete-cohort identity. The README gives the commands and dependencies for these reductions and identifies the additional state/transform and physical-response records required for broader reproduction.